\documentclass[aps,prd,notitlepage,showpacs,nofootinbib,superscriptaddress]{revtex4-1}
\pdfoutput=1
\usepackage[usenames,dvipsnames]{color}  %%%%%%% usenames,dvipsnames required to get BrickRed
\usepackage{graphicx} %%%%%% do not use \usepackage[dvips]{graphicx}%%%% it will creat a clash with color n brickred will not appear%%%%%

\usepackage{bm}
\usepackage{bbold}
\usepackage{latexsym}
\usepackage{amsthm}
\usepackage{amsmath}
\usepackage{wrapfig}
\usepackage{amssymb}
\usepackage{cancel}
\usepackage{multirow}%\usepackage{dsfont}
\usepackage{pst-3d}                  % PSTricks 3D framed boxes
\usepackage{pst-grad}                % PSTricks gradient mode
\usepackage{pst-node}                % PSTricks nodes
\usepackage{pst-slpe}                % Improved PSTricks gradients
\usepackage{pst-plot}                %
\usepackage{pst-coil}                %
\usepackage{pst-math}                %
\usepackage{pstricks-add}            %
\usepackage{rotating}
\usepackage{geometry}
\usepackage{hhline}
\usepackage{amsmath}
\usepackage{float}
\usepackage{fancybox}
\usepackage[utf8]{inputenc}

\usepackage[colorlinks=true,citecolor=darkred,urlcolor=darkred, pdfborder={0 0 0}]{hyperref}
\usepackage[normalem]{ulem}

\usepackage{braket}
\definecolor{linkcolor}{rgb}{0,0,0.5}
\allowdisplaybreaks
\allowdisplaybreaks[1]
\allowdisplaybreaks[2]

\definecolor{greenLinks}{rgb}{0, 0.6, 0}
\definecolor{blueLinks}{rgb}{0, 0, 0.6}
\definecolor{redLinks}{rgb}{0.6, 0, 0}
\definecolor{tempText}{rgb}{0.55, 0.10,0.67}
\definecolor{eprintLinks}{rgb}{0.4, 0.4, 0.4}
\definecolor{journalLinks}{rgb}{0.6, 0, 0}
\newcommand {\ignore}[1]{}

\usepackage{soul}
\definecolor{mightnightblue}{RGB}{25,25,112}

\definecolor{brown}{rgb}{0.59, 0.29, 0.0}

\definecolor{darkred}{rgb}{0.6,0,0}

\def\lfv{lepton flavour violation }

\def\lsim{\mathrel{\rlap{\lower4pt\hbox{\hskip1pt$\sim$}}
    \raise1pt\hbox{$<$}}}
\def\gsim{\mathrel{\rlap{\lower4pt\hbox{\hskip1pt$\sim$}}
    \raise1pt\hbox{$>$}}}

\def\U1s{$\mathrm{U_{1}^{(a)}\otimes U_{1}^{(b)}\otimes U_{1}^{(c)}\otimes U_{1}^{(d)}\otimes U_{1}^{(e)}}$ }
\def\3211{$\mathrm{SU(3) \otimes SU(2)_L \otimes U(1)_R \otimes U(1)_{B-L}}$ }
\def\321{$\mathrm{SU(3) \otimes SU(2) \otimes U(1)}$ }
\def\422{$\mathrm{SU(4) \otimes SU(2) \otimes SU(2)_R}$ }

\newcommand{\AddrAHEP}{%
  AHEP Group, Institut de F\'{i}sica Corpuscular --
  C.S.I.C./Universitat de Val\`{e}ncia, Parc Cient\'ific de Paterna.\\
 C/ Catedr\'atico Jos\'e Beltr\'an, 2 E-46980 Paterna (Valencia), SPAIN}
 
 \newcommand{\AddrMPI}{Max-Planck-Institut f\"ur Kernphysik, Saupfercheckweg 1, 69117 Heidelberg, GERMANY}

\begin{document}

\title{\boldmath \color{BrickRed} 
Natural axion model from flavour}

\author{Salvador Centelles Chuli\'{a}}\email{salcen@ific.uv.es}
\affiliation{\AddrAHEP}
\author{Christian D\"oring}\email{christian.doering@mpi-hd.mpg.de}
\affiliation{\AddrMPI}
\author{Werner Rodejohann}\email{werner.rodejohann@mpi-hd.mpg.de}
\affiliation{\AddrMPI}
\author{Ulises J. Salda\~na-Salazar}\email{ulises.saldana@mpi-hd.mpg.de}
\affiliation{\AddrMPI}

\begin{abstract}
\noindent
We explore a common symmetrical origin for two long standing problems in particle physics: the strong CP and the fermion mass hierarchy problems. The Peccei--Quinn mechanism solves the former one with an anomalous global $U(1)_{\rm PQ}$ symmetry. Here we investigate how this $U(1)_{\rm PQ}$ could at the same time explain the fermion mass hierarchy. We work in the context of a four-Higgs-doublet model which explains all quark and charged fermion masses with natural, i.e.\ order $1$, Yukawa couplings. Moreover, the axion of the model constitutes a viable dark matter candidate and neutrino masses are incorporated via the standard type-I seesaw mechanism. A simple extension of the model allows for Dirac neutrinos. 
\end{abstract}

\maketitle

\section{Introduction}
\label{sec:introduction}
\noindent
Though having been very successful, the standard model of particle physics (SM) lacks the explanation for a number of experimental observations that strongly point towards, amongst other things, the existence of dark matter (DM) and a non-zero neutrino mass. 
Furthermore the SM poses so called fine-tuning problems in which symmetry-allowed parameters have to be assigned very small numbers in order to fit the experimental results. One of such problems is the strong CP problem. The non-trivial topological structure of the QCD vacuum leads to a CP-violating total derivative term $\theta\, G_{\mu\nu}\tilde{G}^{\mu\nu}$ in the QCD Lagrangian with the gluon field strength tensor $G_{\mu\nu}$, its dual,  and the CP violating angle $\theta$. Naturally\footnote{Typically, in the literature, the term 'natural' has two distinct meanings. In the first one a small parameter is said to be natural when setting it to zero leads to a bigger symmetry of the Lagrangian, i.e.\ it has a symmetry protection \cite{tHooft:1979rat}. In turn, this implies that renormalization corrections to this parameter will be proportional to itself.
Examples of this behaviour in the SM include the strong CP-violating parameter, protected by CP symmetry, and the small fermionic Yukawas, protected by chiral symmetry. In the second meaning \cite{Dirac:1938mt}, which we use throughout this work, we say that a dimensionless parameter is natural when it is of $\mathcal{O} (1)$. An abnormally small parameter 
may be a hint towards a hidden physical mechanism that explains its smallness.}
 one expects  such parameters to be of $\mathcal{O} (1)$. Measurements of the neutron electric dipole moment, however,  limit the angle from above to be $|\theta|<10^{-10}$, thus hinting to some new physics that explains such smallness.

Among other proposals, the introduction of a light pseudo-scalar field \cite{PhysRevLett.38.1440,PhysRevD.16.1791}, called axion \cite{PhysRevLett.40.279} is the most popular solution to the problem. Hereby $\theta$ is promoted from being a parameter to a dynamical complex pseudo-scalar field, coming along with a QCD anomalous global $U(1)_{\textrm{PQ}}$ symmetry, called Peccei--Quinn (PQ) symmetry. 

The breaking of this symmetry at high scales of order $(10^9-10^{15})\textrm{ GeV}$ leads to a pseudo Goldstone boson, the axion $a(x)$, and in consequence to the presence of the CP-violating term  $a\, G_{\mu\nu}\tilde{G}^{\mu\nu}$. Hence $\theta$ is replaced by the dynamical field $a$ and its vacuum expectation value (VEV) can be chosen to zero, $\langle a\rangle=0$, such that the CP violating term vanishes, and thus dynamically explains the smallness of the value.

Axions are typically produced in the early Universe via non-thermal mechanisms such as vacuum realignment \cite{Dine:1982ah, Preskill:1982cy, Abbott:1982af} or string decay \cite{Davis:1985pt, Davis:1986xc, Vilenkin:1986ku, Harari:1987ht, Davis:1989nj, Lyth:1991bb, Battye:1993jv, Battye:1994au, Nagasawa:1994qu, Chang:1998tb, Yamaguchi:1998gx, Hagmann:2000ja}. These mechanisms lead to a high population of non-relativistic axions. Together with their weak interactions, besides solving the strong CP-problem, axions fulfill the criteria for cold DM. One of the two important scales of the cosmological history of axion DM is the temperature at which the PQ symmetry breaks. One has to distinguish whether the scale breaks during or after inflation, which leads to different allowed mass parameter ranges for the axion. The second important scale is the QCD scale, around which the axion acquires a mass due to non-perturbative effects.

The allowed ranges for the axion mass, in order to have DM, depend strongly on the production mechanism and whether the PQ symmetry breaks before or after inflation. Taking all mechanisms into account opens a wide allowed range for $m_a\le 10^{-2}\,\textrm{eV}$ \cite{DiLuzio:2020wdo}. Experimental studies like ADAMX \cite{Du:2018uak} and CAST \cite{Anastassopoulos:2017ftl} have already released limits and further searches are on going or planned for the future, e.g. ABRACADABRA \cite{Kahn:2016aff}, KLASH \cite{Alesini:2017ifp} or MADMAX \cite{Majorovits:2017ppy}.

Axion models fall into two major categories, the KSVZ \cite{PhysRevLett.43.103,SHIFMAN1980493} and the DFSZ \cite{DINE1981199,Zhitnitsky:1980tq} model. KSVZ models require the introduction of new vectorlike quarks while in DFSZ models the Higgs sector is extended leaving the fermion sector untouched. The model presented in this paper  falls into the DFSZ category by introducing additional Higgs doublets. For more details of the PQ mechanism see the reviews \cite{Srednicki:1985xd, diCortona:2015ldu, DiLuzio:2017pfr, Hook:2018dlk, DiLuzio:2020wdo}.

Another fine tuning problem that the SM faces appears in the fermionic Yukawa sector. From a naturalness point of view, one expects all fermions (i.e.\ the three generations of up-type quarks $q_u$, down-type quarks $q_d$, charged leptons $\ell$ and neutrinos $\nu$) to have masses of the order of the electroweak (EW) scale ($\Lambda_\text{EW} = 174$ GeV) if Yukawas are of $\mathcal{O} (1)$. However, when experimentally measured, various puzzling and apparently unrelated patterns appear, in particular:
\begin{enumerate}
    \item[\textit{i)}] $m_t \sim \Lambda_\text{EW} \gg m_\alpha \quad (\alpha=b,\tau,c,s,\mu,d,u,e)$; 
    \item[\textit{ii)}] $\{m_b, m_\tau,m_c\} \sim 10^{-2} \,\Lambda_\text{EW}$; 
    \item[\textit{iii)}] $\{m_\mu, m_s\} \sim 10^{-3} \,\Lambda_\text{EW}$; 
    \item[\textit{iv)}] $\{m_d, m_u,m_e\} \sim 10^{-5} \,\Lambda_\text{EW}$. 
\end{enumerate}
When enlisted in this way, the patterns seem to suggest four different mass scales. Adding small active neutrino masses adds another one.

The pattern can either be understood by Yukawas of different order and a fixed VEV or by Yukawas of similar order and different order VEVs.
If there is a Higgs doublet for each set of fermions in items $i)-iv)$ above, natural (order $1$) Yukawa couplings are possible when the associated Higgs doublets feature a sharp VEV hierarchy. The focus of attention is then shifted from a hierarchy in the Yukawa sector to a hierarchy in the VEVs of the doublets.

In Ref.~\cite{Rodejohann:2019izm} this idea was realized. 
The original particle content contained in particular 
an enlarged scalar sector with four Higgs doublets and an additional $\mathbb{Z}_2^\prime \times \mathbb{Z}_2^{\prime \prime} \times \mathbb{Z}_2^{\prime \prime \prime}$ symmetry. It is then by the sequential soft-breaking of the aforementioned symmetry that four different mass scales originate. To overcome the appearance of tree-level flavour-changing-neutral-currents (FCNCs) the singular alignment ansatz as described in Ref.~\cite{Rodejohann:2019izm} was implemented. Neutrino masses, which by itself correspond to yet another scale in addition to the ones in $i)-iv)$, were generated via a type-I seesaw mechanism in which the mass scale of set $iv)$ was identified with the Dirac mass matrix. 

Interestingly, these two ideas can be combined into one unified symmetric explanation for both problems. In other words, the axion is not only a viable solution to the strong CP and the DM problems, but it may also be employed as a solution to the fermion mass hierarchy problem, see~\cite{Davidson:1982, Wilczek:1982rv,Davidson:1983fy,Berezhiani:1989fp,Ema:2016ops,Calibbi:2016hwq, Ballesteros:2016xej,Arias-Aragon:2017ewwDavidi:2018sii,Gupta:2019ueh}. 

Here we focus on the 4 Higgs Doublet Model (4HDM) of \cite{Rodejohann:2019izm}. Note that the Abelian nature of the symmetrical group $\mathbb{Z}_2^3$ could naturally originate from a global $U(1)$ symmetry, suggesting a connection to the  origin of the axion. This observation motivates us to try to find a common symmetrical origin for the previously discussed problems (strong CP, DM, and mass hierarchy). In this paper, we explore under what conditions we could make such a statement. Note that, compared to Refs.~\cite{Ema:2016ops,Calibbi:2016hwq, Arias-Aragon:2017eww}, where the Froggatt--Nielsen mechanism~\cite{Froggatt:1978nt} plays the dominant role as a solution to the fermion mass hierarchy puzzle, we instead make use of a renormalizable mechanism for the same target. \\

The paper is organised as follows. In Sec.~\ref{sec:4HDM} a 4HDM in the spirit of \cite{Rodejohann:2019izm} is introduced, in which each fermion set $i)-iv)$ 
couples to only one Higgs doublet. This feature is enforced thanks to a global $U(1)$ symmetry. We describe how this global $U(1)$ is spontaneously broken by two scalar singlets $A$ and $\chi$, which along with the spontaneous electroweak symmetry breaking generates the required VEV hierarchy in a natural way. In Sec.~\ref{sec:PQ} we identify the $U(1)$ global anomalous symmetry with the PQ symmetry and flesh out the phenomenological details of the axion of our model. In particular, we compute the coupling between the axion and photons and show that our model has a non-universal coupling of axions and charged fermions. This can imply the existence of axion-mediated FCNCs which is a potential experimental signal from the model. However, the axion-mediated FCNCs can also be taken to zero with a judicious choice of parameters.

A benchmark numerical scenario showing the main features of the model and satisfying the main experimental constraints has been relegated to Appendix \ref{app:scenario}. Finally, in Appendix~\ref{app:5higgs} we describe an alternative scenario with Dirac neutrinos.

%%%%%%%%%%%%%%%%%%%%%
\section{A 4 Higgs Doublet Model with Order One Yukawa Couplings}
\label{sec:4HDM}
\subsection{General Structure of the Model\label{sec:mod}}
%%%%%%%%%%%%%%%%%%%%%%
\noindent
The central idea of the 4HDM model in \cite{Rodejohann:2019izm} is to explain the mass hierarchy of the fermions by dividing them into four mass classes and assigning each of them to a single Higgs doublet:
\begin{itemize}
    \item $(\{m_t\};\phi_t)$, 
    \item $(\{m_b,m_{\tau},m_c\};\phi_b)$, 
    \item $(\{m_{\mu},m_s\};\phi_{\mu})$, 
    \item $(\{m_d,m_u,m_e\};\phi_d)$,
\end{itemize}
where Yukawa couplings between a Higgs and its assigned fermions are of $\mathcal{O}(1)$ and we define the scalar doublet components as 
\begin{align}\label{eq:higgsdef}
    \phi_\alpha = 
        \begin{pmatrix}
            \phi_\alpha^+ \\    
            (v_\alpha + \frac{\rho_\alpha}{\sqrt{2}})e^{\frac{i a_\alpha}{\sqrt{2}v_\alpha}}
        \end{pmatrix} 
        ,  \qquad (\alpha = t, b, \mu ,d) ,
\end{align}
with $v_t^2 + v_b^2 + v_\mu^2 + v_d^2 = (174 \text{ GeV})^2$. 
The four Higgs doublets must have a sharp VEV hierarchy, which in return gives different order of  magnitude masses to each of the four sets of fermions. It is possible to subsequently induce the VEVs from the next-highest VEV, thereby explaining their hierarchy. 
 This VEV hierarchy can be achieved with a careful choice of $U(1)$ charges and field inventory, as it can be seen in  Tab.~\ref{tab:charges}. Note that the fermion content of the model is exactly the SM, and only the scalar sector has been enlarged.

\begin{table}[t]
\begin{center}
\begin{tabular}{| c || c | c | c |}
  \hline
&   \hspace{0.1cm}  Fields     \hspace{0.1cm}       &  \hspace{0.4cm}  $SU(2)_L \times U(1)_Y$     \hspace{0.4cm}    & \hspace{0.4cm}   $U(1)_{\rm PQ}$            \hspace{0.4cm}                              \\
\hline \hline
\multirow{4}{*}{ \begin{turn}{90} \hspace{-0.6 cm} Leptons \end{turn} } 
&   $L_i$        	     &   ($\mathbf{2}, {-1/2}$)      &   $l$ \\
&   $\nu_{R, i}$         &   ($\mathbf{1}, {0}$)         &   $0$        \\
&   $e_{R}$              &   ($\mathbf{1}, {-1}$)        &   $2l$       \\
&   $\mu_R$              &   ($\mathbf{1}, {-1}$)        &   $2l-k$      \\
&   $\tau_R$    	     &   ($\mathbf{1}, {-1}$)        &   $2l-2k$        \\
\hline \hline                                                         
\multirow{4}{*}{ \begin{turn}{90} \hspace{-1.3 cm} Quarks \end{turn} }      
&   $Q_i$        	     &   ($\mathbf{2}, {1/6}$)      &   $q$         \\
&   $u_{R}$              &   ($\mathbf{1}, {2/3}$)      &   $q-l$        \\
&   $c_R$                &   ($\mathbf{1}, {2/3}$)      &   $q-l+2k$         \\
&   $t_R$                &   ($\mathbf{1}, {2/3}$)      &   $q-l+3k$          \\
&   $d_R$    	         &   ($\mathbf{1}, {-1/3}$)     &   $q+l$           \\
&   $s_R$    	         &   ($\mathbf{1}, {-1/3}$)     &   $q+l-k$           \\
&   $b_R$    	         &   ($\mathbf{1}, {-1/3}$)     &   $q+l-2k$             \\
\hline \hline                                                                             
\multirow{5}{*}{ \begin{turn}{90} \hspace{-0.7 cm} Scalars \end{turn} }                          
& $\phi_t$  	       	 &  ($\mathbf{2}, {1/2}$)        & $3k-l$                   \\
& $\phi_b$  	       	 &  ($\mathbf{2}, {1/2}$)        & $2k-l$                \\
& $\phi_\mu$  	       	 &  ($\mathbf{2}, {1/2}$)        & $k-l$                \\
& $\phi_d$  	       	 &  ($\mathbf{2}, {1/2}$)        & $-l$                 \\
& $\chi$            	 &  ($\mathbf{1}, {0}$)          & $k$                        \\
& $A$                    &  ($\mathbf{1}, {0}$)          & $k/2$                       \\
    \hline
  \end{tabular}
\end{center}
\caption{\footnotesize Particle content and symmetry transformations under the EW gauge group and $U(1)_{\rm PQ}$ symmetry. All the scalars are color singlets while only the quarks transform non-trivially under $SU(3)_C$. The Lagrangian and charges have been checked using \texttt{Sym2Int}~\cite{Fonseca:2017lem}. 
}
 \label{tab:charges}
\end{table}

The presence of an additional scalar singlet ($\chi$), besides the PQ breaking one ($A$), is justified as follows. Once $\phi_t$ acquires a VEV from spontaneous EW-symmetry breaking, the VEVs of the other Higgs doublets are sequentially induced thanks to a cubic interaction with the singlet scalar $\chi$, triggering a cascade in which each Higgs field induces a VEV for the next lower hierarchical level, $v_t\hookrightarrow v_b(v_t)\hookrightarrow v_{\mu}(v_b)\hookrightarrow v_{d}(v_{\mu})$, such that $v_t^2\gg v_{b}^2 \gg v_{\mu}^2\gg v_d^2$. Unless there is extreme tuning, the VEV of $\chi$ is required to be not too far away from the weak scale, which in turn implies that its  Goldstone boson associated to the broken global symmetry cannot be the axion. Instead, another scalar $A$ needs to be introduced, which will be associated to the axion. It also couples to $\chi$, and in fact induces a VEV to it.

\subsection{Fermion Sector\label{sec:fer}}
\noindent
We now give the Lagrangian which satisfies the desired properties and in parallel derive the PQ charges, already shown in Tab.~\ref{tab:charges}, that generate it. Starting from the Yukawa sector of the model we have, for the up and down quarks
\begin{eqnarray}
    -\mathcal{L}_u = & Y_{u, i} \, \overline{Q}_i \, \tilde \phi_d \, u_R \, + \, Y_{c, i} \, \overline{Q}_i \, \tilde \phi_b \, c_R \, + \, Y_{t, i} \, \overline{Q}_i \, \tilde \phi_t \, t_R  \, + \, \text{h.c.},\\
    -\mathcal{L}_d = & Y_{d, i} \, \overline{Q}_i \, \phi_d \, d_R \, + \, Y_{s, i} \, \overline{Q}_i \, \phi_\mu \, s_R \, + \, Y_{b, i} \, \overline{Q}_i \, \phi_b \, b_R  \, + \, \text{h.c.},
\end{eqnarray}
where $i=1,2,3$ sums over the three generations of left-handed quark doublets $Q$.
Note how each quark only receives a single contribution to its mass coming from one Higgs doublet. In the following, we denote the $U(1)$-global charges by $\mathcal{X}$ such that, e.g.\ $\mathcal{X} (Q_i) \equiv \mathcal{X}_{Q_i}$. From this Yukawa Lagrangrian we can immediately find some initial conditions for the PQ charges:
\begin{equation}
    \mathcal{X}_{Q_1} = \mathcal{X}_{Q_2} = \mathcal{X}_{Q_3} = q ,
\end{equation}
while for the right-handed quarks we obtain
\begin{align}
\begin{split}
\mathcal{X}_{u_R} & =  q+\mathcal{X}_{\phi_d} \;, \hspace{1cm}  \mathcal{X}_{d_R} = q-\mathcal{X}_{\phi_d} ,\\
\mathcal{X}_{c_R} & =  q+ \mathcal{X}_{\phi_b}\;, \hspace{1cm}  \mathcal{X}_{s_R} = q-\mathcal{X}_{\phi_\mu} ,\\
\mathcal{X}_{t_R} & =  q+\mathcal{X}_{\phi_t}\;, \hspace{1cm}  \mathcal{X}_{b_R} = q-\mathcal{X}_{\phi_b} .
\end{split}
\end{align}
Analogously, the charged lepton Yukawa couplings are given by
\begin{equation}
    -\mathcal{L}_e = Y_{e, i} \, \overline{L}_i \, \phi_d \, e_R \, + \, Y_{\mu, i} \, \overline{L}_i \, \phi_\mu \, \mu_R \, + \, Y_{\tau, i} \, \overline{L}_i \, \phi_b \, \tau_R  \, + \, \text{h.c.},
    \label{eq:chargedlag}
\end{equation}
which leads to the following conditions for the PQ charges:
\begin{equation}
\begin{gathered}
    \mathcal{X}_{L_1} = \mathcal{X}_{L_2} = \mathcal{X}_{L_3} = l ,\\
\mathcal{X}_{e_R}  =  l-\mathcal{X}_{\phi_d}, \hspace{1cm}  \mathcal{X}_{\mu_R} = l-\mathcal{X}_{\phi_\mu} , \hspace{1cm}
\mathcal{X}_{\tau_R}  =  l-\mathcal{X}_{\phi_b}.
\end{gathered}
\end{equation}
The pieces of the charged Yukawa Lagrangian can be written in matrix form, which explicitly shows the wanted features of the model:
\begin{eqnarray}
\label{eq:matform}
\mathcal{L}_{Q} \,& =& \left(\begin{matrix}  \overline{Q}_1 & \overline{Q}_2 & \overline{Q}_3 \end{matrix} \right)
\left(\begin{matrix}
Y_{u,1} \, \tilde \phi_d & Y_{c,1} \, \tilde \phi_b & Y_{t,1} \, \tilde \phi_t  \\
Y_{u,2} \, \tilde \phi_d & Y_{c,2} \, \tilde \phi_b & Y_{t,2} \, \tilde \phi_t  \\
Y_{u,3} \, \tilde \phi_d & Y_{c,3} \, \tilde \phi_b & Y_{t,3} \, \tilde \phi_t  \\
\end{matrix} \right)
\left(\begin{matrix}  u_R \\ c_R \\ t_R \end{matrix} \right) \, + \, 
\left(\begin{matrix}  \overline{Q}_1 & \overline{Q}_2 & \overline{Q}_3 \end{matrix} \right)
\left(\begin{matrix}
Y_{d,1} \, \phi_d & Y_{s,1} \, \phi_\mu & Y_{b,1} \, \phi_b  \\
Y_{d,2} \, \phi_d & Y_{s,2} \, \phi_\mu & Y_{b,2} \, \phi_b  \\
Y_{d,3} \, \phi_d & Y_{s,3} \, \phi_\mu & Y_{b,3} \, \phi_b  \\
\end{matrix} \right)
\left(\begin{matrix}  d_R \\ s_R \\ b_R \end{matrix} \right)\, + \, \text{h.c.} \nonumber \\
-\mathcal{L}_{e} \, & = &
\left(\begin{matrix}  \overline{L}_1 & \overline{L}_2 & \overline{L}_3 \end{matrix} \right)
\left(\begin{matrix}
Y_{e,1} \, \phi_d & Y_{\mu,1} \, \phi_\mu & Y_{\tau,1} \, \phi_b  \\
Y_{e,2} \, \phi_d & Y_{\mu,2} \, \phi_\mu & Y_{\tau,2} \, \phi_b  \\
Y_{e,3} \, \phi_d & Y_{\mu,3} \, \phi_\mu & Y_{\tau,3} \, \phi_b  \\
\end{matrix} \right)
\left(\begin{matrix}  e_R \\ \mu_R \\ \tau_R \end{matrix} \right) \, + \, \text{h.c.} ,
\end{eqnarray}
where $\mathcal{L}_{Q} =  -(\mathcal{L}_{u} + \mathcal{L}_d)$.

In the neutrino sector, we can easily implement a Majorana type-I seesaw \cite{Minkowski:1977sc,Mohapatra:1979ia,Yanagida:1979as,GellMann:1980vs,Schechter:1981cv} and choose the Dirac term to be given by the VEV of $\phi_d$: 
\begin{equation}
    -\mathcal{L}_\nu = Y_{\nu, ij} \, \overline{L}_i \, \tilde \phi_d \, \nu_{R, j} \, +  \, \frac{M_{ij}}{2} \overline{\nu}_{R, i}^c \, \nu_{R, j}\, + \, \text{h.c.}
    \label{eq:neutrinolag}
\end{equation}
Here $v_d$ is the smallest VEV of the model. Eq.~\eqref{eq:neutrinolag} implies for the PQ charges: 
\begin{equation}
\mathcal{X}_{\nu_{R,1}} = \mathcal{X}_{\nu_{R,2}} = \mathcal{X}_{\nu_{R,3}} = 0 \qquad \text{and} \qquad  \mathcal{X}_{\phi_d} = -l .
\label{eq:chargephid}
\end{equation}
In return, by virtue of the seesaw formula, the light neutrino mass matrix is
\begin{equation}
    {\bf m}_\nu \, = \, -  v_d^2 \, { \bf Y}_\nu {\bf M}^{-1} {\bf Y}_\nu^T  .
\end{equation}
In the 'standard' seesaw case, if we take the Yukawas to be of order $1$ and the neutrino mass scale to be $0.1$ eV then the mediator has to have a mass of $M\sim 10^{14}$ GeV. However, in our model the neutrino masses are given by $v_d\sim 0.001$ GeV instead of the SM VEV. In this case,  the mass of the mediator is given by $M\sim 10^4$ GeV $= 10$ TeV. As shown in \cite{Rodejohann:2019izm} these values still escape the experimental constraints coming from colliders and \lfv processes~\cite{Atre:2009rg,Deppisch:2015qwa}. Let us now note that by relaxing the Yukawas to be of order $0.1$ would lead to a mediator mass of just $100$ GeV. Therefore, this model has a range for observable \lfv and collider physics in certain regions of the parameter space. Notice further that the identification of Yukawa interactions between neutrinos and $\phi_d$ is not required and could have been done with any of the other three scalar doublets without any further consequence for the main features of the model. A detailed type I-seesaw phenomenological study is out of the scope of this work and has been studied extensively in the literature. \\

As a side note, we remark that if the right-handed neutrino charges are taken to be different from zero then the second term in Eq.~\eqref{eq:neutrinolag} would be forbidden and thus neutrinos would be Dirac particles. Their small mass could then be explained by a fifth Higgs doublet whose VEV should be much smaller than the other four, in the same spirit as the explanation to the charged fermion hierarchy. See Appendix~\ref{app:5higgs} for more details. \\

The total Lagrangian made of Eqs.\ (\ref{eq:matform}) and (\ref{eq:neutrinolag})
is the same fermionic Lagrangian as the one shown in \cite{Rodejohann:2019izm} i.e.\ each fermion set receives its mass by the Higgs associated to it. That is, $\{ t \}$, $\{b, \, \tau, \, c\}$, $\{\mu, s\}$ and $\{d, \, u, \, e\}$ acquire their mass by $\phi_t$, $\phi_b$, $\phi_\mu$, $\phi_d$, respectively. As we will show below, the mass hierarchy of such sets is explained by a hierarchy in the VEVs of the different Higgs doublets and order $1$ Yukawa couplings. In particular, and as an example, all masses can be accommodated with order $1$ Yukawas if the VEVs are taken to be $v_t = 174$ GeV, $v_b \sim 1$ GeV, $v_\mu \sim 0.1$ GeV, and $v_d \sim 0.001$ GeV.  A typical problem arising in models with multiple Higgs doublets is the presence of tree-level flavour changing neutral currents. A solution was suggested in \cite{Rodejohann:2019izm} by introducing what was called singular alignment (see Ref.~\cite{Penuelas:2017ikk} for the same ansatz under a different approach). Moreover, this alignment allows us to set aside the problem of mixing as it is  automatically decoupled from the problem of mass. This can be understood as all individual Yukawa matrices are simultaneously diagonalised when moving to the mass basis. We then assume that this mechanism is realized here as well. The next step is to illustrate how the $U(1)$ model acquires the necessary VEV hierarchy, and how the new scalar singlets $\chi$ and $A$ help in that, and how they are connected to the usual axion physics. 

\subsection{Scalar Sector\label{sec:sca}}
\noindent
We now turn our attention to the scalar sector of the model. For the time being, let us study it with the four Higgs doublets and only $\chi$ out of the two singlet scalars. The scalar Lagrangian can be divided into four different pieces, 
\begin{equation}
    {\cal L}_\text{scalar} \supset 
    \mathcal{L}_{S1} + \mathcal{L}_{S2} + \mathcal{L}_{S3} +  \mathcal{L}_\text{kin}.
\end{equation}
Here $\mathcal{L}_\text{kin}$ contains the corresponding kinetic terms of all  scalar fields. The first term groups all the quadratic and quartic terms where a given field and its complex conjugate always appear together: 
\begin{eqnarray}
\label{eq:abelian}
   -\mathcal{L}_{S1} & = &\sum_{a \in \{ t, b, \mu, d\}} \left[\mu_a^2 \, \phi_a^\dagger \, \phi_a \, + \, \frac{\lambda_a}{2} (\phi_a^\dagger \, \phi_a)^2 \right] \, +\, \mu_\chi^2 \chi^\dagger \chi \, + \, \frac{\lambda_{\chi}}{2} (\chi^\dagger \chi)^2 \, + \, \sum_{a \in \{ t, b, \mu, d\}} \lambda_{a\chi} \, (\phi_a^\dagger \, \phi_a)  \chi^\dagger \chi  \nonumber \\
   & & \, + \sum_{a, b \in \{t, b, \mu, d\}; \, a > b} \,\left[ \lambda_{ab1} (\phi_a^\dagger \, \phi_a)(\phi_b^\dagger \, \phi_b) \, +  \, \lambda_{ab2} (\phi_a^\dagger \, \phi_b)(\phi_b^\dagger \, \phi_a)\right] .  \\ \nonumber
\end{eqnarray}
We note that $\mathcal{L}_{S1}$ has an accidental $U(1)^5$ symmetry, one $U(1)$ per scalar field. If the scalar Lagrangian were just formed by $\mathcal{L}_{S1}$ then the model would have five Goldstone bosons.
However, we add the following terms to our Lagrangian in order to reduce the symmetry of the model $U(1)^5 \rightarrow U(1)_Y \times U(1)_{\rm PQ}$ and therefore, by construction, only two Goldstone bosons remain. These extra terms are given by
\begin{eqnarray} \label{eq:lagcascade0}
&& -\mathcal{L}_{S2} = \, \lambda_{tbb\mu} (\phi_t^\dagger \phi_b) (\phi_\mu^\dagger \phi_b) \, + \, \lambda_{tb\mu d1} (\phi_t^\dagger \phi_b) (\phi_d^\dagger \phi_\mu) \, + \, \lambda_{tb\mu d2} (\phi_t^\dagger \phi_\mu) (\phi_d^\dagger \phi_b)  \, + \, \lambda_{b \mu\mu d} (\phi_b^\dagger \phi_\mu) (\phi_d^\dagger \phi_\mu) \, + \, \text{h.c.}, \\
\label{eq:lagcascade}&& -\mathcal{L}_{S3} = -\, \kappa_{tb\chi} \, (\phi_t^\dagger \phi_b) \chi \, - \,\kappa_{b\mu\chi} \, (\phi_b^\dagger \phi_\mu) \chi \, - \,\kappa_{\mu d\chi} \, (\phi_\mu^\dagger \phi_d) \chi \, + \lambda_{t\mu\chi \chi} \, (\phi_t^\dagger \phi_\mu) \chi^2 \, + \,\lambda_{b d\chi \chi} \, (\phi_b^\dagger \phi_d) \chi^2 \, + \, \text{h.c.},
\end{eqnarray}
where we have assumed a CP symmetrical scalar potential by considering real couplings. Now, note that $\mathcal{L}_{S_2}$ breaks $U(1)^5 \rightarrow U(1)_Y \times U(1)_{\rm PQ} \times U(1)_\chi$ and $\mathcal{L}_{S3}$ breaks it further into $U(1)_Y \times U(1)_{\rm PQ}$. While $\mathcal{L}_{S2}$ has no further consequences of interest to us, it will be automatically generated if $\mathcal{L}_{S3}$ is allowed by the symmetries of the model. On the other hand, the dimensionful couplings of $\mathcal{L}_{S3}$ are the relevant sources of the `VEV cascade' and thus key to our construction, as we show later. 

 After spontaneous EW symmetry breaking it can be shown that the mass matrix for the CP-even scalars has rank $5$, while the mass matrix for CP-odd scalars has rank $3$, i.e.\ it has two massless eigenstates, the two Goldstone bosons mentioned before. We come back to this point in Sec.~\ref{sec:PQ}.
 
 Note that in order to allow the $\kappa$ terms in $\mathcal{L}_{S3}$, the charges of two consecutive Higgs doublets must be separated by a constant charge which is at the same time the charge of $\chi$. With this in mind, and with Eq.~\eqref{eq:chargephid}, we find the last conditions for our PQ charges:
\begin{eqnarray}
\label{eq:chargesphi}
&\mathcal{X}_{\chi} = k, \hspace{1cm}  \mathcal{X}_{\phi_d} = -l,&\\
&\mathcal{X}_{\phi_\mu} = k-l, \hspace{1cm}  \mathcal{X}_{\phi_b} = 2k-l, \hspace{1cm}  \mathcal{X}_{\phi_t} = 3k-l& .
\end{eqnarray}
We can now express all the PQ charges in terms of $q$, $l$, and $k$, which are the charges of the quark doublet, the lepton doublet, and the singlet scalar $\chi$, respectively. This information is already shown in Tab.~\ref{tab:charges}. 

Let us now focus on the term $\mathcal{L}_{S3}$ in Eq.~\eqref{eq:lagcascade} and assume that 
\begin{equation}
    \mu_t^2 < 0, \qquad
    \mu_b^2 > 0, \qquad
    \mu_\mu^2 > 0 , \quad \text{and} \quad
    \mu_d^2 >0 ,
\end{equation}   
such that, in the absence of Eqs.~\eqref{eq:lagcascade0} and  \eqref{eq:lagcascade}, only $\phi_t$ develops a VEV, 
$v_t \neq 0$, while all others remain zero. Moreover, the singlet scalar $\chi$ also has a VEV, for which the axion is responsible (as discussed below). Under these features, Eq.~\eqref{eq:lagcascade} induces small and hierarchical VEVs, as obtained from the minimum conditions 
\begin{equation} 
    \frac{\partial ({\cal L}_{S1} + {\cal L}_{S2} + {\cal L}_{S3})}{\partial \rho_\alpha} \Bigg|_\text{min} = 0 , \qquad (\alpha = t,b,\mu,d) , 
\end{equation} 
where $\rho_\alpha$ is defined in Eq.~\eqref{eq:higgsdef}, and from which one correspondingly obtains an intricate system of polynomial equations:
\begin{align}
\begin{split}
 & v_t \left[\mu_t^2 + \lambda_t v_t^2 + \lambda_{t\chi} v_\chi^2 + \sum_{k=b,\mu,d}\widetilde{\lambda}_{tk}v_k^2 \right] + \lambda_{tbb\mu}v_\mu v_b^2 + \widetilde{\lambda}_{tb\mu d} v_b v_\mu v_d + v_\chi (\lambda_{t\mu \chi \chi}v_\mu - v_b \kappa_{tb\chi}) = 0 , \\
 & v_b \left[\mu_b^2 + \lambda_b v_b^2 + \lambda_{b\chi} v_\chi^2 + \sum_{k=t,\mu,d} \widetilde{\lambda}_{bk} v_k^2 \right] + \lambda_{b\mu\mu d}v_\mu^2 v_d + 2 \lambda_{tbb\mu} v_t v_b v_\mu - (\kappa_{tb\chi} v_t + \kappa_{b\mu\chi} v_\mu) v_\chi + \lambda_{bd\chi\chi}v_d v_\chi^2 = 0 , \\
 & v_\mu \left[\mu_\mu^2 + \lambda_\mu v_\mu^2 + \lambda_{\mu\chi} v_\chi^2 + \sum_{k=t,b,d}\widetilde{\lambda}_{k\mu}v_k^2 \right] + v_t (\lambda_{tbb\mu} v_b^2  + \lambda_{t\mu\chi\chi} v_\chi^2) +\widetilde{\lambda}_{tb\mu d} v_t v_b v_d + 2 \lambda_{b\mu\mu d}v_b v_\mu v_d - (\kappa_{b\mu\chi} v_b+\kappa_{\mu d\chi} v_d) v_\chi = 0, \\
 & v_d \left[\mu_d^2 + \lambda_d v_d^2 + \lambda_{d\chi} v_\chi^2 + \sum_{k=t,b,\mu}\widetilde{\lambda}_{kd}v_k^2 \right] + \lambda_{b\mu\mu d} v_b v_\mu^2 + \lambda_{bd\chi\chi} v_b v_\chi^2 + \widetilde{\lambda}_{tb\mu d} v_t v_b v_\mu - \kappa_{\mu d \chi} v_\mu v_\chi = 0,
\end{split}
\end{align}
where, in general, for arbitrary values there is no guaranteed solution and we have defined $\widetilde{\lambda}_x = \lambda_{x1} + \lambda_{x2}$. We find that by imposing the conditions $\kappa_{ij\chi} v_\chi < ({\cal O}( 10 - 100 ) \text{ GeV})^2$ and all  dimensionless couplings appearing in Eqs.~\eqref{eq:lagcascade0} and~\eqref{eq:lagcascade} to be smaller than $10^{-1}$, we guarantee that each Higgs field induces a VEV for the next lower hierarchical level. Furthermore, both conditions make also possible to directly apply the results of Ref.~\cite{Rodejohann:2019izm} to our
case, see Appendix~\ref{app:scenario} for further details.
In return, the VEVs can be nicely approximated by
\begin{equation} \label{eq:VEVs}
    \begin{gathered}
    v_t \simeq \sqrt{\frac{-\mu_t^2}{\lambda_t}} , \qquad
    v_b \simeq \frac{\kappa_{tb\chi} v_\chi v_t}{\mu_b^2 + (\lambda_{tb1}+\lambda_{tb2}) v_t^2+ \lambda_{b\chi} v_\chi^2} , \\
    v_\mu \simeq \frac{\kappa_{b\mu\chi}v_\chi v_b}{\mu_\mu^2 + (\lambda_{t\mu 1}+\lambda_{t\mu 2}) v_t^2 + \lambda_{\mu\chi} v_\chi^2} , \qquad
    v_d \simeq \frac{\kappa_{\mu d\chi}v_\chi v_\mu}{\mu_d^2 + (\lambda_{td 1} + \lambda_{td 2}) v_t^2 + \lambda_{d\chi} v_\chi^2} .
    \end{gathered}
\end{equation}
Addition of the second singlet scalar, whose CP-odd component is to be later identified with the dominant contribution to the axion, brings about the last piece of information for the full scalar Lagrangian, $ {\cal L}_\text{scalar} = 
    \mathcal{L}_{S1} + \mathcal{L}_{S2} + \mathcal{L}_{S3} + \mathcal{L}_A + \mathcal{L}_\text{kin}$: 
\begin{eqnarray} \notag
    -{\cal L}_{A_1}& = &\mu_A^2 A^\dagger A \, + \, \frac{\lambda_{A}}{2} (A^\dagger A)^2 
    \, + \, \sum_{a \in \{ t, b, \mu, d\}} \left[ \lambda_{aA} \, (\phi_a^\dagger \, \phi_a)  A^\dagger A \right]
    \, + \, \lambda_{\chi A}  (\chi^\dagger \chi) \, (A^\dagger A) \\
  \notag  -{\cal L}_{A_2} & = & \, - \,\kappa_{A A \chi} (A A \chi^\dagger \, + \,\lambda_{t b A} \phi_t^\dagger \phi_b A^2\, + \,\lambda_{b \mu A} \,\phi_b^\dagger \phi_\mu A^2\textbf{} +\textbf{} \lambda_{\mu d A}\textbf{} \phi_\mu^\dagger \phi_d A^2\textbf{} +\, \text{h.c.} \\
    {\cal L}_A &= &{\cal L}_{A_1} + {\cal L}_{A_2}.
\end{eqnarray}
A major consequence of the previous equation\footnote{With the introduction of a sixth scalar, the previous discussion on accidental symmetries changes as follows: The self-conjugated scalar terms $\mathcal{L}_{S_1} \, + \, \mathcal{L}_{A_1}$ have an accidental $U(1)^6$ symmetry. The presence of the non-Hermitian terms $\mathcal{L}_{S2} + \mathcal{L}_{S3} + \mathcal{L}_{A_2}$ explicitly breaks it,  $U(1)^6 \rightarrow U(1)_Y \times U(1)_{\rm PQ}$.}  is that, through the minimization conditions
\begin{equation}
    \frac{\partial {\cal L}_\text{scalar}}{\partial \text{Re}(A) } \Bigg|_\text{min} = 0  \qquad \text{and}\qquad
    \frac{\partial {\cal L}_\text{scalar}}{\partial\text{Re}(\chi) } \Bigg|_\text{min} = 0 ,
\end{equation}
which imply
\begin{equation}
\begin{gathered}
 v_A (-2 v_\chi \kappa_{AA\chi} + \lambda_A v_A^2 + \lambda_{\chi A} v_\chi^2 +\mu_A^2) =  0 ,\\
 v_\chi\left(\mu_\chi^2 + \lambda_\chi v_\chi^2 +  \lambda_{\chi A} v_A^2 + \sum_{k=t,b,\mu,d}\lambda_{k\chi} v_k^2 \right) -\sum_{j=tb,b\mu,\mu d} \kappa_{j\chi} v_j^2 -\kappa_{AA\chi} v_A^2  + 2 (\lambda_{bd\chi\chi}v_b v_d + \lambda_{t\mu\chi\chi}v_t v_\mu) v_\chi  = 0  ,
\end{gathered}
\end{equation}
it shows how $A$ induces a small VEV to $\chi$; 
\begin{equation} \label{eq:minChi}
v_A \simeq \sqrt{-\frac{\mu_A^2 }{\lambda_A}} \qquad \text{and} \qquad 
    v_\chi \simeq \frac{\kappa_{AA\chi} v_A^2}{\mu_\chi^2 + \lambda_{\chi A}v_A^2} . 
\end{equation}
Here we have assumed $\mu_A^2 < 0$ and $\mu^2_\chi >0$. Realize the induced hierarchical nature of the VEVs, $v_A \gg v_\chi$. For example, for $v_A \sim 10^{12} \text{ GeV}$ and $\mu_\chi \sim 10^7 \text{ GeV}$, we obtain $v_\chi \simeq \kappa_{AA\chi} = {\cal O}(1 - 10) \text{ GeV}$. Moreover, to avoid large loop corrections and PQ-scale contributions to the scalar masses and VEVs we fine-tune the mixing between the singlet scalar $A$ and the doublet scalar fields such that it is negligible. This is the well known hierarchy problem which in the DFSZ model always appears, as well as in any model with a scale much higher than the EW. \\

As a conclusion for this section, let us point out the importance of the global symmetry $U(1)_{\rm PQ}$. Thanks to the symmetry breaking pattern of the scalars one can naturally obtain a sharp VEV hierarchy between the four Higgs doublets. At the same time, the $U(1)$ structure of the fermions combined with this sharp VEV hierarchy leads to the observed fermion mass hierarchy with order $1$ Yukawas. The whole mechanism is summarised in Figure~\ref{fig:scheme} and an explicit realization is offered in Appendix~\ref{app:scenario}. Note that no predictions for the fermion mixings can be made, since we are making use of an Abelian symmetry and assuming singularly aligned Yukawa matrices.

In what follows, we study the consequences of identifying the global symmetry $U(1)$ with the PQ symmetry and the Goldstone boson with the axion, which is the main point of this paper. 
\begin{figure}[t]
    \centering
        \includegraphics[width=0.7\textwidth]{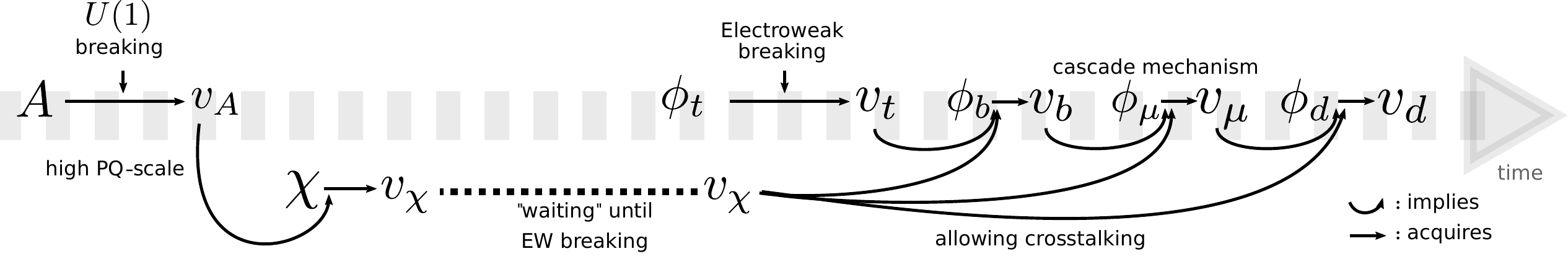}
    \caption{Schematic view of the development of the radial field components of the model. At high scales, the new global $U(1)_{\rm PQ}$ symmetry spontaneously breaks. The radial component (solid line) of $A$ then acquires a VEV while the phase becomes a massless Goldstone boson (GB), i.e.\ the axion (not shown here). The radial component, however, induces a VEV for the radial component of $\chi$. Until EW symmetry breaking this has no impact whatsoever, but as soon as the radial component of $\phi_t$ acquires a VEV by spontaneous EW symmetry breaking, a cascade is triggered. Together, $v_{\chi}$ and $v_t$ imply a VEV for $v_b$ and so on. Due to the scalar terms, explicitly breaking $U(1)^6 \rightarrow U(1)_Y \times U(1)_{\rm PQ}$, there are only two massless GBs at the end of the chain, instead of six,  which are a mixture of the CP-odd components of all the scalar fields.  }
    \label{fig:scheme}
\end{figure}

\section{The Peccei--Quinn sector}
\label{sec:PQ}
\noindent
Having introduced our model and explained the cascade mechanism, this section is dedicated to the phase of the PQ-breaking field $A$, which we identify with the axion. We start by explicitly deriving the conserved current from the PQ-symmetry using Noether's theorem which is given by
    \begin{equation}
    \label{eq:scacurrent}
        j^\mu_{\rm PQ} = \frac{\partial \mathcal{L}}{\partial({\partial_\mu \psi_i})} \delta \psi_i(x) + \frac{\partial \mathcal{L}}{\partial({\partial_\mu \varphi_j})} \delta \varphi_j(x) \rightarrow  j^\mu_{\rm PQ} = i \alpha \left( \mathcal{X}_{\psi_i} \bar{\psi}_i \gamma^\mu \psi_i \, + \, \varphi_j^\dagger \mathcal{X}_{\varphi_j} \partial^\mu \varphi_j \right) \, + \, \text{ h.c.},
    \end{equation}
    where $\mathcal{L}$ is the full Lagrangian, $\psi_i$ and $\varphi_i$ run over all the fermion and scalar fields, respectively, and $\mathcal{X}_{\psi_i}$ and $\mathcal{X}_{\varphi_i}$ are the PQ charges of the field in the subscript; $\delta \psi_i = i \alpha \mathcal{X}_{\psi_i} \psi_i$, $\delta \varphi_i = i \alpha \mathcal{X}_{\varphi_i} \psi_i$, $\alpha$ is the infinitesimal parameter of the continuous transformation and a sum over the indices $i$ and $j$ is understood.
    By definition only the phases of the Higgs fields and the PQ breaking scalar contribute to the axion. Hence we will now drop the fermionic fields, the charged Higgs and the radial components of the neutral Higgses in $j^\mu_{\rm PQ}$. Then we have 
   \begin{eqnarray}
     \phi_k \supset	v_k e^{i \frac{a_k}{\sqrt{2}\, v_k}} \left(\begin{matrix} 0 \\ 1\end{matrix}\right), \hspace{0.5 cm} (k\in \{t,b,\mu,d\})     , \hspace{2cm} \chi \supset	v_\chi e^{i \frac{a_\chi}{\sqrt{2}\, v_\chi}} , \hspace{2cm} A \supset	v_A e^{i \frac{a_A}{\sqrt{2}\, v_A}} ,
     \end{eqnarray}
which upon substitution in Eq.~\eqref{eq:scacurrent} immediately yields
     \begin{equation}
         \label{eq:axioncurrent}
         j^\mu_{\rm PQ}|_a =  v_i \mathcal{X}_i \partial^\mu a_i ,
     \end{equation}
    where we have removed a constant global factor $-\sqrt{2} \alpha$. Now, note that by defining
    \begin{equation}
        a = \frac{1}{v_a} \sum_i \mathcal{X}_i v_i a_i \qquad \qquad
        \text{and} \qquad \qquad v_a^2 = \sum_i \mathcal{X}_i^2 v_i^2 \approx \mathcal{X}_A^2 v_A^2 ,
    \end{equation}
    we obtain $j^\mu_{\rm PQ}|_a = v_a \partial^\mu a$ and $\langle 0| j^\mu_{\rm PQ}|_a |a\rangle = i v_a p^\mu$, as expected for a massless Goldstone boson and thus $a$ can be identified with the axion.

This kind of models for the axion fall under the category of DFSZ models. Specifically, since we use four Higgs doublets, it would fall under the category of DFSZ-IV~\cite{DiLuzio:2017pfr}.  
This category considers the maximum allowed number of Higgs doublets, $n_H$, contributing to the ratio between the QCD and QED anomalies, which is $n_H = 9$ (one \textit{per} charged fermion). Hence, our model is the most minimal realization of a DFSZ-IV theory where each flavour, of a given fermion species, couples to a different scalar doublet whose VEV judiciously represents one of four groups of similar masses. We will now flesh out the phenomenological implications of the axion of our model. \\

In order to avoid a dangerous kinetic mixing between the axion and the $Z$ boson, see  Ref.\ \cite{DiLuzio:2020wdo}, we must impose the orthogonality between the PQ and the hypercharge currents, i.e.\ 
    \begin{equation}
    \label{eq:orthocurrents}
        j^\mu_{\rm PQ}|_a \, \, j^{Y}_\mu|_a = 0 \qquad \Rightarrow \qquad \sum_{i \in \{t,b,\mu, d, \chi, A\}} Y_i \mathcal{X}_i v_i^2 = 0 \;,
    \end{equation}
   with $i$ running for all the scalars in the model. Since the singlets $\chi$ and $A$ do not carry hypercharge and all doublets have the same hypercharge,  Eq.~\eqref{eq:orthocurrents} simplifies to
        \begin{equation}
        \sum_{i \in \{t,b,\mu, d\}} \mathcal{X}_i v_i^2 = 0 \qquad \Rightarrow \qquad  (3k-l) \,v_t^2 + (2k-l) \,v_b^2 + (k-l)\, v_\mu^2 -l \, v_d^2 = 0 \;,
     \end{equation}
    with $i$ running only for the four scalar doublets. Thanks to the sharp hierarchy between the VEVs the equation can be solved by taking $k \simeq l/3$, since  $v_t$ dominates\footnote{For a more general scenario, where the four VEVs are arbitrary, the orthogonality condition would require:
    \begin{equation}
        \mathcal{X}_t = s_{\beta_t}^2 s_{\beta_b}^2 s_{\beta_\mu}^2 c_{\beta_b}^2 c_{\beta_\mu}^2 , \qquad
        \mathcal{X}_b = s_{\beta_b}^2 s_{\beta_\mu}^2 c_{\beta_t}^2 c_{\beta_\mu}^2 , \qquad
        \mathcal{X}_\mu = -s_{\beta_\mu}^2 c_{\beta_t}^2 c_{\beta_b}^2 ,\qquad
        \mathcal{X}_d = -c_{\beta_t}^2 c_{\beta_b}^2 c_{\beta_\mu}^2 \;,
    \end{equation}
    where the angles are defined by 
    \begin{align}
    \sin \beta_t  = \sqrt{\frac{v_b^2+v_\mu^2+v_d^2}{v_t^2+v_b^2+v_\mu^2+v_d^2}} \;, \qquad
    \sin \beta_b  = \sqrt{\frac{v_\mu^2+v_d^2}{v_b^2+v_\mu^2+v_d^2}}  \;,\qquad
    \sin \beta_\mu  = \sqrt{\frac{v_d^2}{v_\mu^2+v_d^2}} \;.
\end{align}
    }. \\

\begin{figure}[t]
    \centering
        \includegraphics[width=0.7\textwidth]{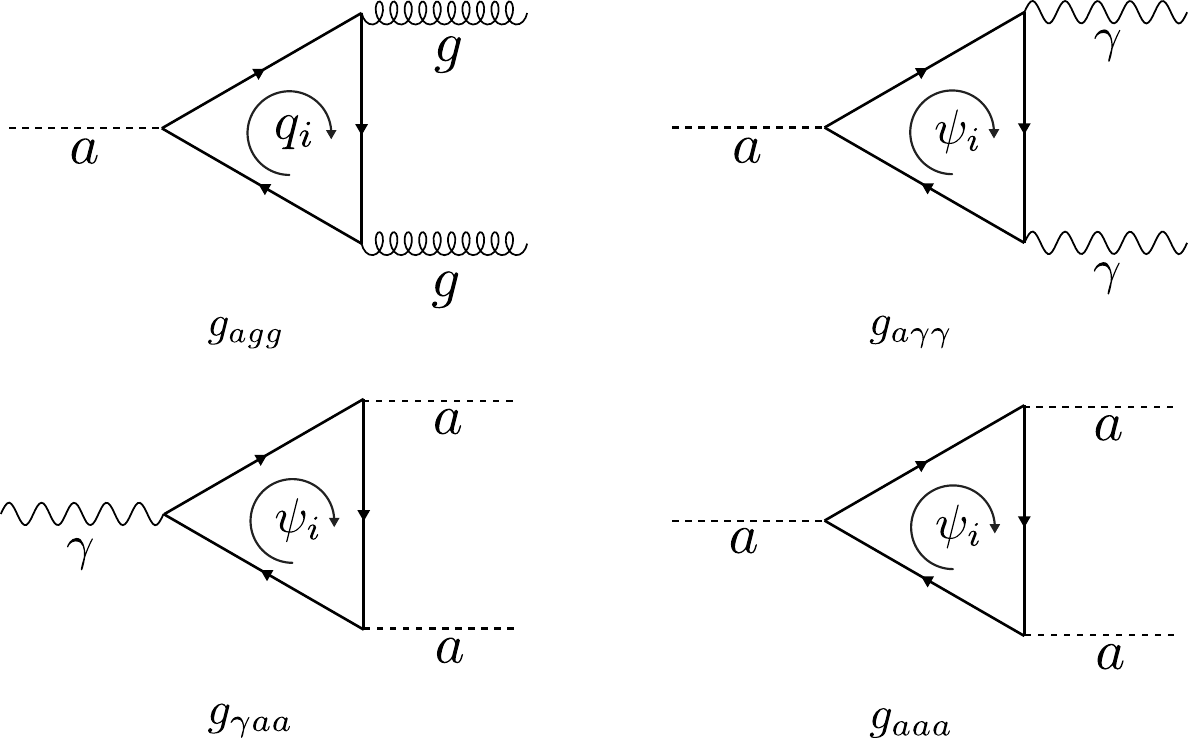}
    \caption{Triangular diagrams associated with the axion in our model. Here $a$, $\gamma$, and $g$ denote the axion, the photon, and the gluon, respectively, while $\psi_i$ denotes all the fermions and $q_i$ the six quarks. The first panel comes from the $SU(3)_C^2 \times U(1)_{\rm PQ}$ anomaly and generates instanton effects implying CP conservation in the strong sector. The second panel comes from the $U(1)_{\rm EM}^2 \times U(1)_{\rm PQ}$ anomaly and allows a decay of an axion into two photons and thus a detection possibility, suppressed by $1/f_a$. The third and fourth panels depicts the vertices $aa\gamma$ and $aaa$ which are suppressed by $1/f_a^2$ and $1/f_a^3$, respectively.} 
    \label{fig:TriagularFeyDiag}
\end{figure}

The choice of charges leads to a number of triangular anomalies.  
First, the fermion $U(1)_{\rm PQ}$ charges given by Tab.~\ref{tab:charges} are anomalous with respect to the colour group. In particular, we have computed the coefficients for all the possible anomalies using \texttt{Susyno} \cite{Fonseca:2011sy}. Since the gauge structure is the same as in the SM and our model does not feature new fermions, the gauge group anomalies are cancelled like in the standard case. However, all the triangular anomalies related to  $U(1)_{\rm PQ}$ and the gauge group will be non-zero. We give special attention to $U(1)_{\rm EM}^2 \times U(1)_{\rm PQ}$, $U(1)_{\rm EM} \times U(1)_{\rm PQ}^2$, and $SU(3)_C^2 \times U(1)_{\rm PQ}$. The non-trivial vacuum of $SU(3)_C$ will generate instanton effects which will lead to CP conservation in the QCD vacuum, like in the usual PQ mechanism. The four vertices generated by these anomalies are shown in Fig.~\ref{fig:TriagularFeyDiag}.

We now compute the ratio between the QCD anomaly and the QED anomaly, typically called $E/N$ in the literature. This ratio is given by
\begin{equation}
    \frac{E}{N} = \frac{2}{3} + 2 \frac{\sum_i{\left( (\mathcal{X}_{e_{L, i}} - \mathcal{X}_{e_{R, i}})+ (\mathcal{X}_{u_{L, i}} - \mathcal{X}_{u_{R, i}})\right)} }{\sum_i\left( (\mathcal{X}_{d_{L, i}} - \mathcal{X}_{d_{R, i}})+ (\mathcal{X}_{u_{L, i}} - \mathcal{X}_{u_{R, i}})\right)} = \frac{8}{3} \;,
\end{equation}
where $i$ runs over the three generations of fermions. The first equality is completely general (as long as the fermion content with non-trivial gauge charges is the same as in the SM) while the second equality is the particular value of our model. This result is irrespective of the choice of the charges $q$, $l$, and $k$. It is a common result in axion models in which, like ours, the $U(1)$ charges in the down quarks and charged leptons imply coupling to the same scalar doublets,
\begin{equation}
    \mathcal{X}_{d_{L, i}} - \mathcal{X}_{d_{R, i}} = \mathcal{X}_{e_{L, i}} - \mathcal{X}_{e_{R, i}} , \qquad (i=1,2,3) \;,
\end{equation} 
although there are other possibilities. For example, a simpler model with just two Higgs doublets and universal PQ charges for each fermion type could have either $E/N = 8/3$ or $E/N = 2/3$, depending on which Higgs couples to the electron. Additionally, it is necessary to remark here that $E/N = 8/3$ is a shared feature amongst axion models where the fermion mass hierarchy is explained through the PQ symmetry group, e.g. see Refs.~\cite{Cheng:1995fd,Barbieri:1995uv,Barbieri:1997tu,Dudas:2013pja,Falkowski:2015zwa,Calibbi:2016hwq,Bjorkeroth:2018ipq}. Explicitly, $E/N=8/3$, is a consequence of any theory where the down-quark and charged lepton mass hierarchy is explained by the same scalars irrespective of the up-quarks, which is very reasonable as  they have similar masses, $m_{d,i} = {\cal O}(1) \, m_{e,i}$ ($i=1,2,3$).

As mentioned before, axions can be a major component of the DM density and serve due to their properties as cold DM\footnote{Thermally produced axions could constitute hot DM \cite{Duffy:2009ig}.}. In order to have the axion to be DM and to determine the amount it contributes, one has to make specific choices. In particular, if the PQ breaking occurs before inflation,  topological defects get inflated away and their contribution to axion production is negligible. In this case, the dominant axion production mechanism is the misalignment mechanism. Then, the axion DM abundance will depend on the initial misalignment angle $\theta_i$. 
However, these issues depend on the phenomenological and cosmological scenarios and the concrete realization of model parameters which go beyond the scope of this work.  Since our model is just a realization of a DFSZ-type of model, we directly apply general results from earlier studies of this class. Hence the only requirement for the axion in our model being cold DM is to have a mass $m_a^{\textrm{DM}}\le 10^{-2}\,\textrm{eV}$ in order to fall in the range of any of the allowed scenarios. However, note that having multi-component DM is also a viable possibility \cite{Cao:2007fy,Bian:2013wna, Aoki:2012ub,Alves:2016bib, Ahmed:2017dbb,Aoki:2017eqn,Herrero-Garcia:2017vrl,Bhattacharya:2018cgx,Poulin:2018kap,YaserAyazi:2018lrv,Aoki:2018gjf} and the axion could be responsible for only a fraction of the DM density, thus relaxing the constraints. For more details on axion DM we refer the reader to \cite{Duffy:2009ig}.

\begin{figure}[t]
    \centering
        \includegraphics[width=0.7\textwidth]{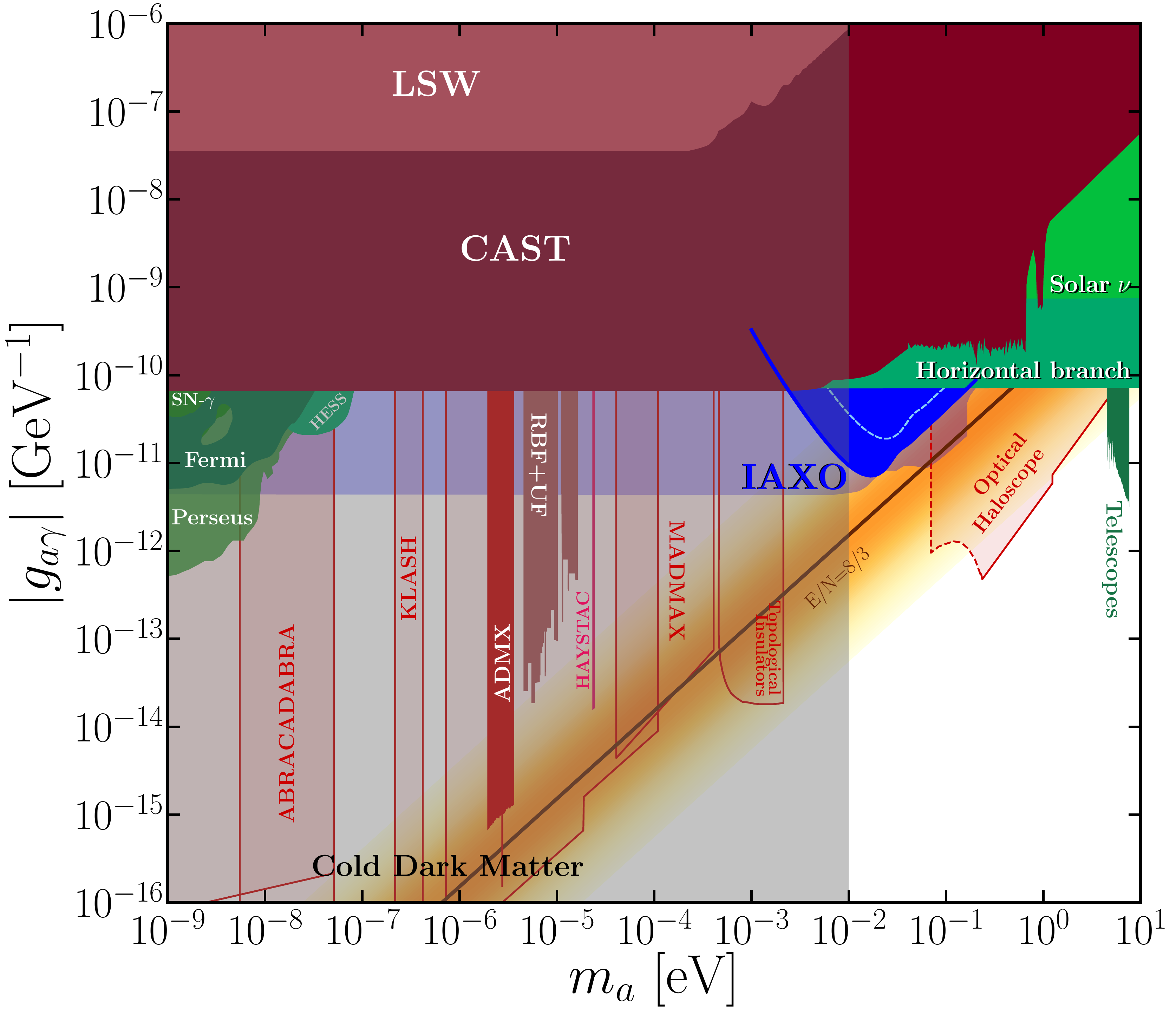}
    \caption{Axion exclusion plot from \cite{Dafni:2018tvj} with the derived $E/N = 8/3$ of this model. Shown is the axion-photon coupling versus the axion mass. The colored areas are excluded regions while transparent colored regions will be tested by future experiments
    \cite{PhysRevD.42.1297,Du:2018uak,Zhong:2018rsr,2013ITAS...23T0604S,Du:2018uak,Majorovits:2017ppy,Kahn:2016aff,Marsh:2018dlj,Alesini:2017ifp,Baryakhtar:2018doz,Payez:2014xsa,Wouters:2013iya,TheFermi-LAT:2016zue,Grin:2006aw,Ayala:2014pea,Vinyoles:2015aba,Anastassopoulos:2017ftl,Ehret:2010mh,Ballou:2015cka,DellaValle:2015xxa}. Also shown in grey the mass region for axion-DM.}
    \label{fig:Limitplot}
\end{figure}

Following \cite{Srednicki:1985xd, diCortona:2015ldu, DiLuzio:2017pfr}, we can compute the mass of the axion in our model
\begin{equation}
    m_a \approx 57.(6)(4) \mu \text{eV} \left( \frac{10^{11} \text{ \text{ GeV}}}{f_a}\right) = m^{\rm DM}_{a} \left( \frac{10^{11} \text{GeV}}{f_a}\right) \;,
\label{eq:amass}
\end{equation}
where we have defined $m^{\rm DM}_{a} = 57.6 \,\mu \text{eV}$ as the highest mass of the axion that leads to a correct one-component DM relic density, assuming the axion is the main component of dark matter, and $f_a = v_a / (2N)$ is the axion decay constant. This happens when $f_a \approx 10^{11} \text{ GeV}$. Remember that the constraint for axion dark matter is $f_a > 10^{11}$ GeV \cite{ediss14919}.  Note that the only model-dependence of this formula is via $f_a$. The effective coupling between the axion and the photon, which will depend on $E/N = 8/3$ in our model, can also be computed and reads 
\begin{equation}
    g_{a\gamma \gamma} \approx \frac{\alpha_{\rm EM} }{2 \pi f_a} \left(\frac{E}{N}-1.92(4) \right) \approx 8.67 \times 10^{-4} \frac{1}{f_a} = g^{\rm DM}_{a\gamma \gamma} \left( \frac{10^{11} \text{ GeV}}{f_a}\right)\;,
    \label{eq:gay}
\end{equation}
where we have defined $g^{\rm DM}_{a\gamma \gamma} = 8.67 \times 10^{-15} \text{ GeV}^{-1}$ as the maximum coupling between the axion and two photons in a model with axion DM and $E/N=8/3$. Combining Eqs.~\eqref{eq:amass} and \eqref{eq:gay} leads to a simple relation between $m_a$ and $g_{a\gamma \gamma}$, which is shown graphically in Fig.~\ref{fig:Limitplot} including the present and future experimental constraints: 
\begin{equation}
    g_{a\gamma\gamma}= \left(\frac{g^{\rm DM}_{a\gamma \gamma}}{m^{\rm DM}_{a}}\right)\, m_a \;.
\end{equation}
Although major parts of the axion DM parameter space with $E/N=8/3$ are still out of reach of current experiments, exciting times are ahead of us, with experiments like MADMAX \cite{Majorovits:2017ppy}, ADMX \cite{Du:2018uak}, and the use of topological insulators \cite{Marsh:2018dlj} having enough sensitivity to probe this region in the future. Also note that this conclusion holds for any axion model with the same ratio of color and electromagnetic anomalies.

Analogously, we could compute the coupling constant of the vertex $a a \gamma$ and the vertex $aaa$, both shown in Fig.~\ref{fig:TriagularFeyDiag}, this time driven by the anomalies  $U(1)_{\rm PQ}^2 \times U(1)_{\rm EM}$ and $U(1)_{\rm PQ}^3$, respectively. However, these processes will be suppressed by $1/f_a^2$ and $1/ f_a^3$, respectively, and thus are expected to be extremely small.\\

As a final remark, let us point out the current situation of our model on FCNCs mediated by the axion. Following the derivation in Ref.~\cite{DiLuzio:2020wdo}, we can write an effective interaction between the axion and fermions in the form
\begin{eqnarray}
\mathcal{L}_{a\psi} = \frac{\partial_\mu a}{2 f_a} \left[  \bar{\psi}_i \gamma^\mu \left(C^V_{\psi ij} - C^A_{\psi ij} \gamma_5\right) \psi_j\right],
\end{eqnarray}
where $\psi_i$ determines the fermion type (up quarks, down quarks, charged leptons, and neutrinos) and a sum over the indices $i,j$, which run from $1$ to $3$, is understood. The coupling constants are given in general by
\begin{eqnarray}
C^{V,A}_{\psi ij} = \frac{1}{2N} \left({\bf U}_L^{\psi \dagger}{\bf X}_{\psi L} {\bf U}_L^\psi \pm {\bf U}_R^{\psi \dagger} {\bf X}_{\psi R} {\bf U}_R^\psi \right)_{ij},
\end{eqnarray}
with $U^\psi_{L(R)}$ are the left (right) unitary diagonalization matrices of each fermion type $\psi$ and $X_{\psi L,R}$ is a diagonal matrix with the PQ charges of the type with the given chirality. Given our choice of $U(1)$ charges, as shown in Table~\ref{tab:charges}, we find that ${\bf X}_{\psi L}$ is a matrix proportional to the identity and thus the equation simplifies to
\begin{eqnarray}
C^{V,A}_{\psi ij} = \frac{1}{2N} \left(\mathcal{X}_{\psi_L} \, {\bf I}  \pm {\bf U}_R^{\psi \dagger} {\bf X}_{\psi R} {\bf U}_R^\psi \right)_{ij}.
\end{eqnarray}
Moreover, realize that 
\begin{equation}
    {\bf X}_{\psi R} = g(q,l) {\bf I} \, \pm \,
    \begin{pmatrix}
    0 & 0 & 0 \\
    0 & \alpha k & 0 \\
    0 & 0 & (\alpha +1) k
    \end{pmatrix} ,
\end{equation}
where $g(q,l) \equiv \{q-l,q+l,l\}$, $\{+,-,-\}$, and $\alpha = \{2,1,1\}$ correspond to the up-quarks, down-quarks, and charged leptons, respectively. Now, notice that it is only $k$ which may induce flavour-violating couplings. Their possible presence, however, can be removed by a judicious choice of ${\bf U}_R^{\psi}$ which is not constrained by fermionic mixing as its counterpart, ${\bf U}_L^{\psi}$. This freedom is a direct a consequence of having applied singular alignment to the model.

\section{Conclusions}
\label{sec:conclusions}
\noindent
We have presented a model where the strong CP, DM, neutrino masses, and fermion mass hierarchy problems find a common symmetrical origin in the Peccei--Quinn global and Abelian symmetry group, $U(1)_{\rm PQ}$. The construction allows all dimensionless couplings in the EW and strong sector to be $\mathcal{O} (1)$. We have required the introduction of two gauge singlet and complex scalars and three Higgs doublets apart from the SM one. Additionally, small neutrino masses find their origin in a standard type-I seesaw mechanism by including three heavy right-handed neutrinos, whose charge under PQ is assumed to be trivial. Relaxing this assumption can lead to Dirac neutrinos, whose small mass can be explained in the same spirit as the charged fermion mass hierarchy by the addition of a fifth Higgs doublet, $\phi_\nu$, with a very small VEV $v_{\phi_\nu} \sim 1 \text{ eV}$. 

To avoid tree-level FCNCs we have applied the singular alignment ansatz in the Yukawa sector. Moreover, since this ansatz provides a linear realization of the principle of minimal flavour violation, the appearance of FCNCs at the loop-level poses no risk. Two singlet scalars are required, as otherwise, the large PQ scale induces unacceptable amounts of fine-tuning in the model. 

The ratio between the colour and electromagnetic anomalies is predicted to be $E/N = 8/3$, irrespective of the explicit $U(1)_{\rm PQ}$ charge assignments of the model. This value originates from allocating the same scalar doublets to couple to the down-quarks and charged leptons. Our approach is the most minimal realization of a DFSZ-IV model wherein all charged fermions couple to a different Higgs doublet. Moreover, non-universal charges for each fermion family leads to the possibility of appearance of axion-mediated FCNCs suppressed by $1/f_a$. However, this FCNCs can be taken to vanish with a judicious choice of unconstrained parameters, thanks to the singular alignment ansatz.

The interplay between Higgs and axion physics plays a major role in distinguishing our model from other axion models. The study of the SM-like Higgs couplings to the other fermions will already be sufficient to test some of our predictions.\\

In conclusion, we present a possible solution to two fine tuning problems in the SM, namely, the flavour and strong CP problems, by combining them in a single unified symmetrical framework with natural dimensionless parameters. In turn, the model is compatible with the parameter space in which the axion is the Dark Matter, thus solving a third issue of the Standard Model.

\begin{acknowledgments}
\noindent
S.C.C.\ would like to thank the Max-Planck-Institut für Kernphysik in Heidelberg for hospitality during his visit, where this work was initiated. 
The work of S.C.C.\ is supported by the Spanish grants SEV-2014-0398, FPA2017-85216-P (AEI/FEDER, UE), PROMETEO/2018/165 (Generalitat  Valenciana), 
Red Consolider MultiDark FPA2017-90566-REDC and BES-2016-076643. The work of W.R.\ is supported by the DFG with grant RO 2516/7-1 in the Heisenberg program.
U.J.S.S.\ acknowledges support from CONACYT (M\'exico).
\end{acknowledgments}

\appendix

\section{Benchmark scenario}
\label{app:scenario}
\noindent
As a proof of principle, we provide an explicit realization of the VEV cascade and show its implications. First of all, notice that when the set of dimensionless couplings 
\begin{equation}
   {\cal C} = \{\lambda_{tbb\mu},\lambda_{tb\mu d1},\lambda_{tb\mu d2},\lambda_{b\mu\mu d},\lambda_{t\mu \chi \chi}, \lambda_{bd\chi \chi}, \lambda_{tA},\lambda_{bA},\lambda_{\mu A},\lambda_{d A},\lambda_{tbA},\lambda_{b\mu A},\lambda_{\mu d A} \}
\end{equation}
is set to zero, ${\cal C} \rightarrow 0$, and after integrating out the  singlet scalars, $\chi$ and $A$, we recover the original model of Ref.~\cite{Rodejohann:2019izm} with all of its couplings of the type $Z_{ab}$ vanishing,
\begin{equation}
\{Z_{tb},Z_{t\mu},Z_{td},Z_{b\mu},Z_{bd},Z_{\mu d} \}= 0 .    
\end{equation}
This allows us to know which part of the parameter space will closely resemble the original model. 

Let us now, step by step, argue the following considerations:
\begin{itemize}
    \item To avoid the hierarchy problem we set to zero those couplings inducing very large mixing between the axion and the Higgs doublets:
    \begin{equation}
        \{ \lambda_{tA},\lambda_{bA},\lambda_{\mu A},\lambda_{d A},\lambda_{tbA},\lambda_{b\mu A},\lambda_{\mu d A} \} =0 .
    \end{equation}
    
    \item To avoid unnecessary interference during the VEV cascade mechanism we set an upper bound to the following couplings:
    \begin{equation}
        \{\lambda_{tbb\mu},\lambda_{tb\mu d1},\lambda_{tb\mu d2},\lambda_{b\mu\mu d},\lambda_{t\mu \chi \chi}, \lambda_{bd\chi \chi} \} \leq 10^{-1} .
    \end{equation}
  One may allow couplings to be order $1$, but the smallest VEV, namely $v_d$, will suffer from non-negligible contributions and thus Eq.~\eqref{eq:VEVs} will not longer be completely valid.
    
    \item Bounding from below the scalar potential requires some necessary conditions which nevertheless are not sufficient:
    \begin{equation}
    \begin{gathered}
        \lambda_{t,b,\mu,d,\chi,A} \geq 0 , 
        \qquad
        \lambda_{tb1} \geq -\sqrt{\lambda_t \lambda_b} ,
        \qquad
        \lambda_{t\mu 1} \geq -\sqrt{\lambda_t \lambda_\mu},
        \qquad
         \lambda_{td1} \geq -\sqrt{\lambda_t \lambda_d},
        \\
         \lambda_{b\mu 1} \geq -\sqrt{\lambda_b \lambda_\mu},
         \qquad
        \lambda_{bd 1} \geq -\sqrt{\lambda_b \lambda_d},
        \qquad
        \lambda_{\mu d 1} \geq -\sqrt{\lambda_\mu \lambda_d},
        \qquad
        \lambda_{\chi A} \geq - \sqrt{\lambda_\chi \lambda_A},
        \\
         \lambda_{t\chi} \geq -\sqrt{\lambda_t \lambda_\chi},
         \qquad
        \lambda_{b\chi} \geq -\sqrt{\lambda_b \lambda_\chi},
        \qquad
        \lambda_{\mu \chi} \geq -\sqrt{\lambda_\mu \lambda_\chi},
        \qquad
        \lambda_{d \chi} \geq - \sqrt{ \lambda_d \lambda_\chi}.
    \end{gathered}
    \end{equation}
    
    \item To ensure unitarity and perturbativity bounds we set the limits
    \begin{equation}
        0 < \lambda_{t,b,\mu,d,\chi,A} \lesssim 2 ,
        \qquad
        -4 \lesssim (\lambda_{ij1} +\lambda_{ij2}) \lesssim 2 ,
        \qquad
        |\lambda_{ij1}| \lesssim 3 ,
        \qquad
        |\lambda_{ij2}| \lesssim 3 ,
    \end{equation}
    where $ij=\{tb,t\mu,td,b\mu,bd,\mu d\}$. These limits were obtained using the K-matrix formalism~\cite{Wigner:1946zz,Wigner:1947zz,Chung:1995dx}.
    
    \item We consider $|\mu_A| \gg |\mu_\chi|$ which is sufficient to guarantee that to a very good degree of approximation $v_\chi \sim \kappa_{AA\chi}$ holds.
    
    \item To create the right hierarchical structure among the EW-VEVs we assume
    \begin{equation}
        \{\kappa_{AA\chi},\kappa_{tb\chi}, \kappa_{b\mu\chi}, \kappa_{\mu d\chi}\} \in {\cal O}(1-100) \text{ GeV} .
    \end{equation}
    However, as the off-diagonal entries of the scalar mass matrices depend on the combination $v_\chi \kappa_{ij\chi} = \kappa_{AA\chi} \kappa_{ij\chi}$ to avoid large mixing, we require these two kind of couplings to behave inversely proportional to each other, $\kappa_{AA\chi} \propto 1/\kappa_{ij\chi}$.
    
     \item To guarantee that the lightest scalar, to be associated to the SM-Higgs, mainly comes from $\phi_t$ we consider the explicit values:
    \begin{equation} \label{eq:mhconditions}
        |\mu_t| = 88.5 \text{ GeV} \qquad \text{ and } \qquad
        \lambda_t = 0.26 ,
    \end{equation}
    which imply $m_h = 125 \text{ GeV}$. Realize that as long as $\phi_t$ is the initial field acquiring a VEV, such that $v_t \gg v_b \gg v_\mu \gg v_d$ and $v_\chi \sim \kappa_{AA\chi}$, then Eq.~\eqref{eq:mhconditions} is the only set of values giving rise to $m_h$, mimicking those already appearing in the SM. That is, there is no ambiguity on which neutral scalar becomes the lightest one. 
    
    \item Any charged scalar mass should obey the constraint~\cite{Abbiendi:2013hk}:
    \begin{equation}
        m_{H^\pm} > 80 \text{ GeV} .
    \end{equation}
    
    \item Measured couplings of fermions with the lightest neutral scalar should lie in the range:
    \begin{equation}
    \kappa_t = 1.02^{+0.19}_{-0.15} , \qquad
    \kappa_b = 0.91^{+0.17}_{-0.16} , \qquad
    \kappa_\tau = 0.93 \pm 0.13 , \qquad
    \kappa_\mu = 0.72^{+0.50}_{-0.72} ,
    \end{equation}
    obtained from combined fits of data taken at $\sqrt{s} = 13 \text{ TeV}$~\cite{Sirunyan:2018koj}. The coupling modifiers, $\kappa_j$, are defined such that for a given production process or decay mode one has $\kappa_j^2 = \sigma_j / \sigma_\text{SM}$ or $\kappa_j^2 = \Gamma_j / \Gamma_\text{SM}$, respectively. This parametrisation allows to conclude that in the SM scenario the coupling modifier is always positive and equal to unity, $\kappa_j = 1$. We merge the bottom and tau lepton coupling modifiers to $\widetilde{\kappa}_b = 0.92^{+0.15}_{-0.14}$ as the bottom quark, tau lepton, and charm quark couple to the same Higgs doublet.
    
    \item We only select those scenarios where the VEVs imply Yukawa couplings (in the mass basis) not larger or equal than one, $y_f \leq 1$. The notion of ${\cal O}(1)$ Yukawa couplings is, in fact, ambiguous. Here we use the range $y_f \in [0.1,1]$, meaning:
    \begin{equation}
        v_t = 174 \text{ GeV} , \quad
        v_b \in [m_b, 10\, m_c] , \quad
        v_\mu \in [m_\mu,10\, m_s] , \quad
        v_d \in [m_d,10\, m_e] .
    \end{equation}
    
    \item The mass difference among the heavy scalars is taken to be less than $50 \text{ GeV}$ to guarantee negligible contributions ($\lesssim 10^{-4}$) to the $\rho$-parameter,
    \begin{equation}
        \Delta \rho = 0.0005 \pm 0.0005 \;(\pm 0.0009) ,
    \end{equation}
    that is, it should be consistent with the maximum allowed deviation from the SM expectation~\cite{Tanabashi:2018oca}. The first and second uncertainty corresponds to the choice of fixing the oblique parameter $U$ to zero or not within the multi-parameter fit. For further details please refer to Ref.~\cite{Rodejohann:2019izm}.
    
    \item For the analysis, we consider the following ranges for the $\mu_{b,\mu,d}$ parameters
    \begin{equation}
        \mu_b \in (125,1000] \text{ GeV} , \qquad
        \mu_\mu \in (125,1000] \text{ GeV}, \qquad
        \mu_d \in (125,1000] \text{ GeV} ,
    \end{equation}
    while for those ones in charge of inducing the cascade mechanism,
    \begin{equation}
        \kappa_{tb\chi} \in [0.5,500] \text{ GeV} , \qquad
        \kappa_{b\mu\chi} \in [0.5,500] \text{ GeV} , \qquad
        \kappa_{\mu d \chi} \in [0.5,500] \text{ GeV} .
    \end{equation}
    
    \item We want to add here that as we are assuming singular alignment, there are no tree-level FCNCs. Furthermore, as already discussed in Ref.~\cite{Rodejohann:2019izm}, the ansatz of singular alignment corresponds to a linear realization of the principle of minimal flavour violation~\cite{Kagan:2009bn}, which also helps to understand that the appearance of FCNCs at loop level will still be sufficiently suppressed respecting all experimental bounds~\cite{Buras:2010mh}. 
    
\end{itemize}

\begin{figure}[t]
    \centering
        \includegraphics[scale=0.34]{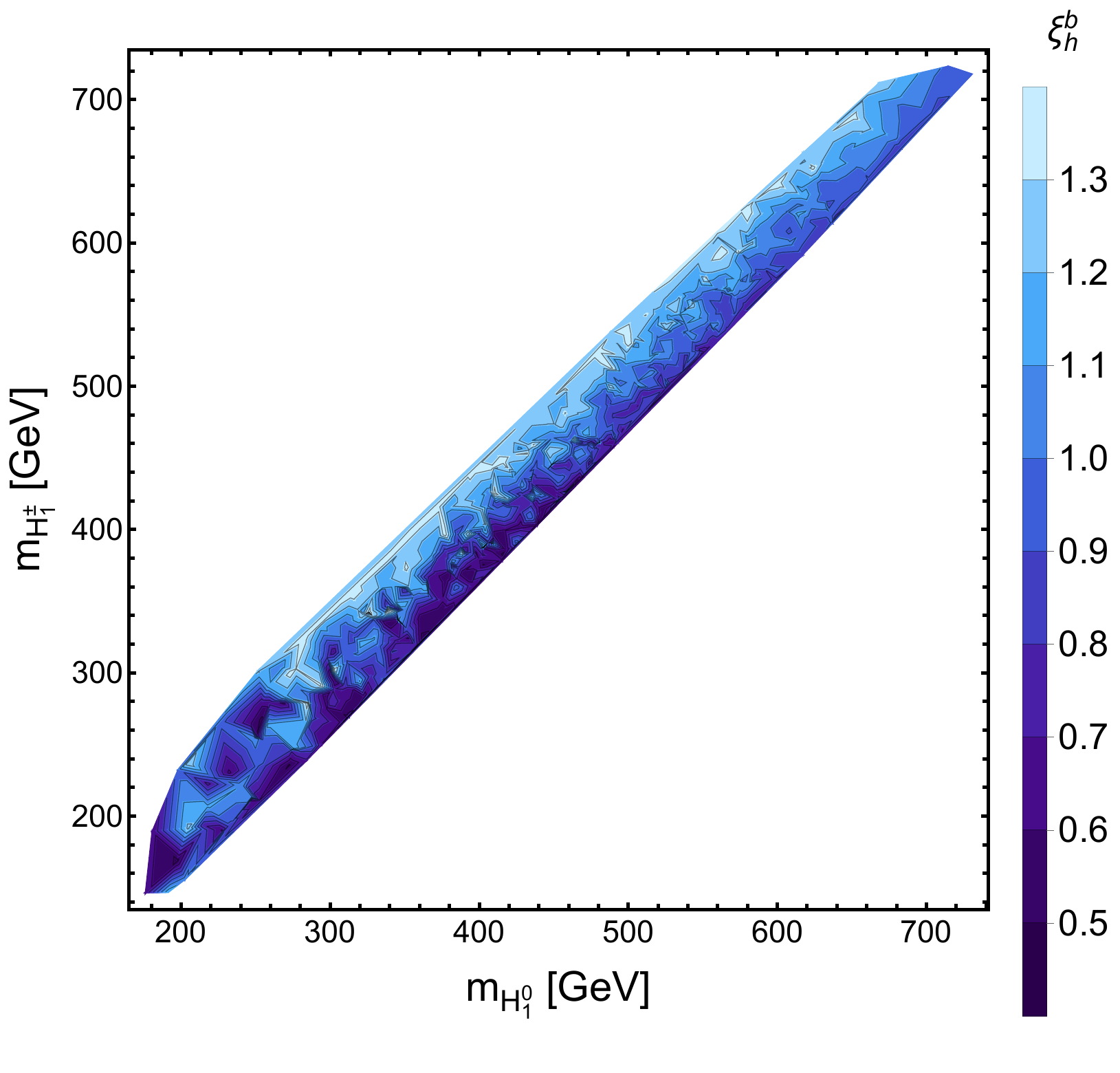}
        \includegraphics[scale=0.34]{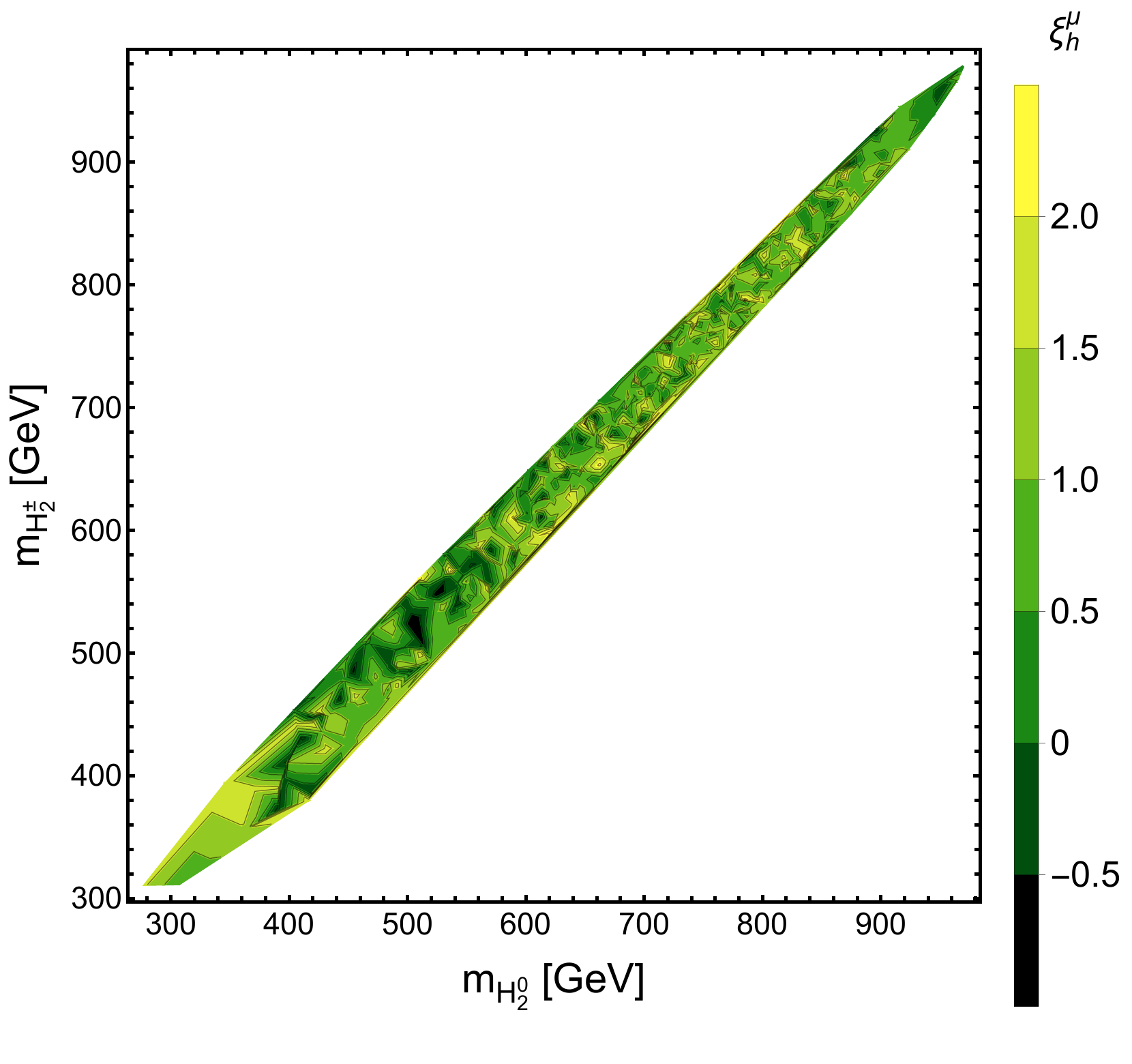}
        \includegraphics[scale=0.34]{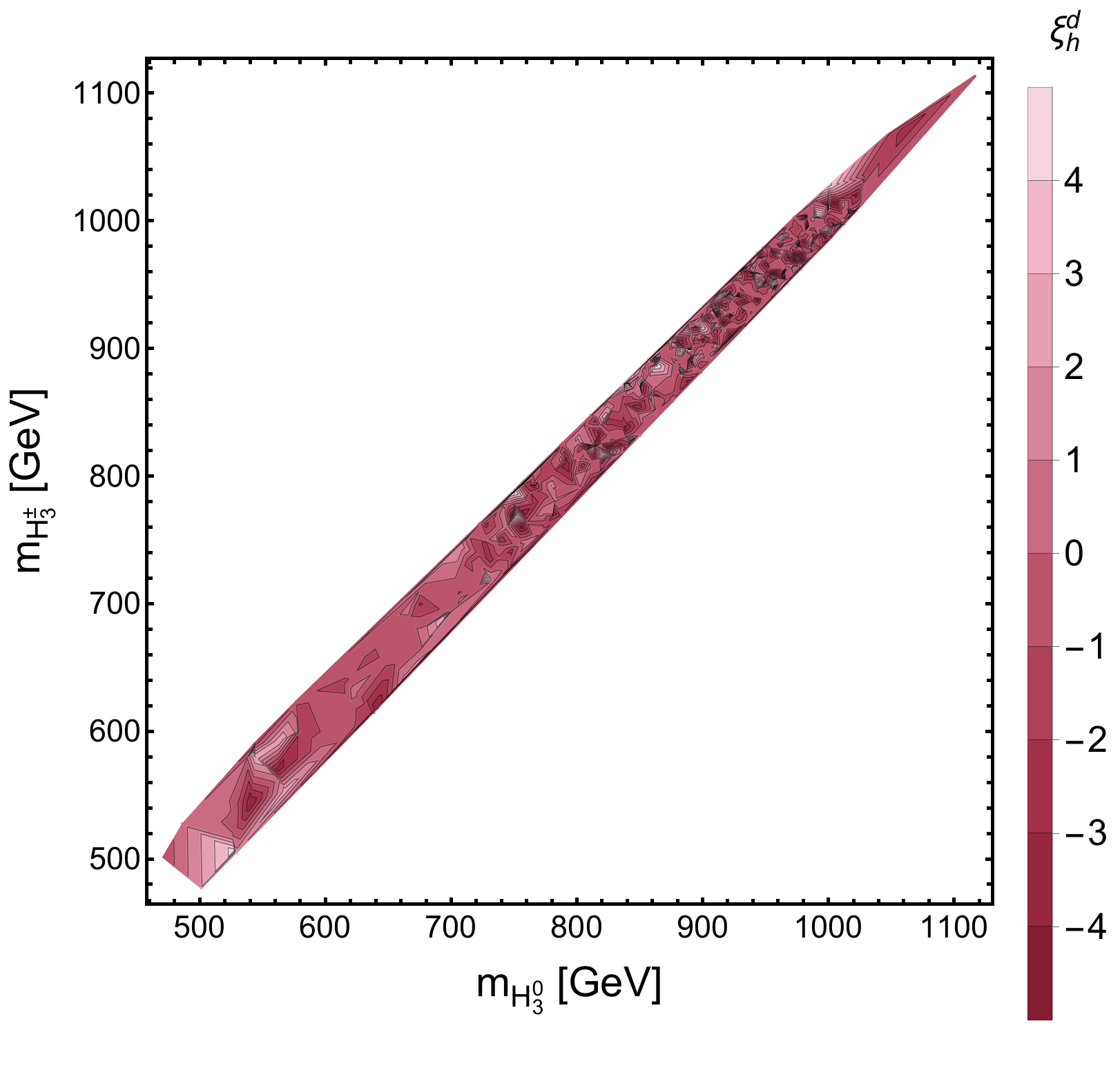}
    \caption{Correlation plots of the scalar masses. In each figure we have contour plotted the value of the effective couplings, in the allowed 3$\sigma$ ranges, between the fermions and the SM-like Higgs. In all cases $\xi^t_h = 1$, $m_h = 125 \text{ GeV}$, and equal plots can be found by replacing $H^0_k \rightarrow A^0_k$. The chaotic behaviour in the couplings can be associated to the large amount of parameters.}
    \label{fig:correlations}
\end{figure}

We denote the scalar mass eigenstates by
\begin{align}\notag
    \text{CP-even neutral scalars:  } &
    \{h,H_1^0,H^0_2,H^0_3\} , \\ \notag
    \text{CP-odd neutral scalars:  } &
    \{A_1^0,A_2^0,A^0_3\} , \\ \notag  
    \text{Electrically charged scalars:  } &
    \{H_1^\pm,H^\pm_2,H^\pm_3\} ,
\end{align}
where we consider $h$ to describe the lightest neutral scalar with SM-like behaviour, $m_h = 125 \text{ GeV}$. On the other hand, we parametrize the deviations from the SM-couplings between the fermions and the SM-like Higgs by $\xi^f_h$, as implied from
\begin{equation} 
    -{\cal L}_y \supset \sum_f \frac{m_f}{(246 \text{ GeV})} \xi^f_h \bar{f}f h ,
\end{equation}
where for $\xi^f_h= 1$ one recovers the SM case. 

\begin{figure}[t]
    \centering
        \includegraphics[scale=0.27]{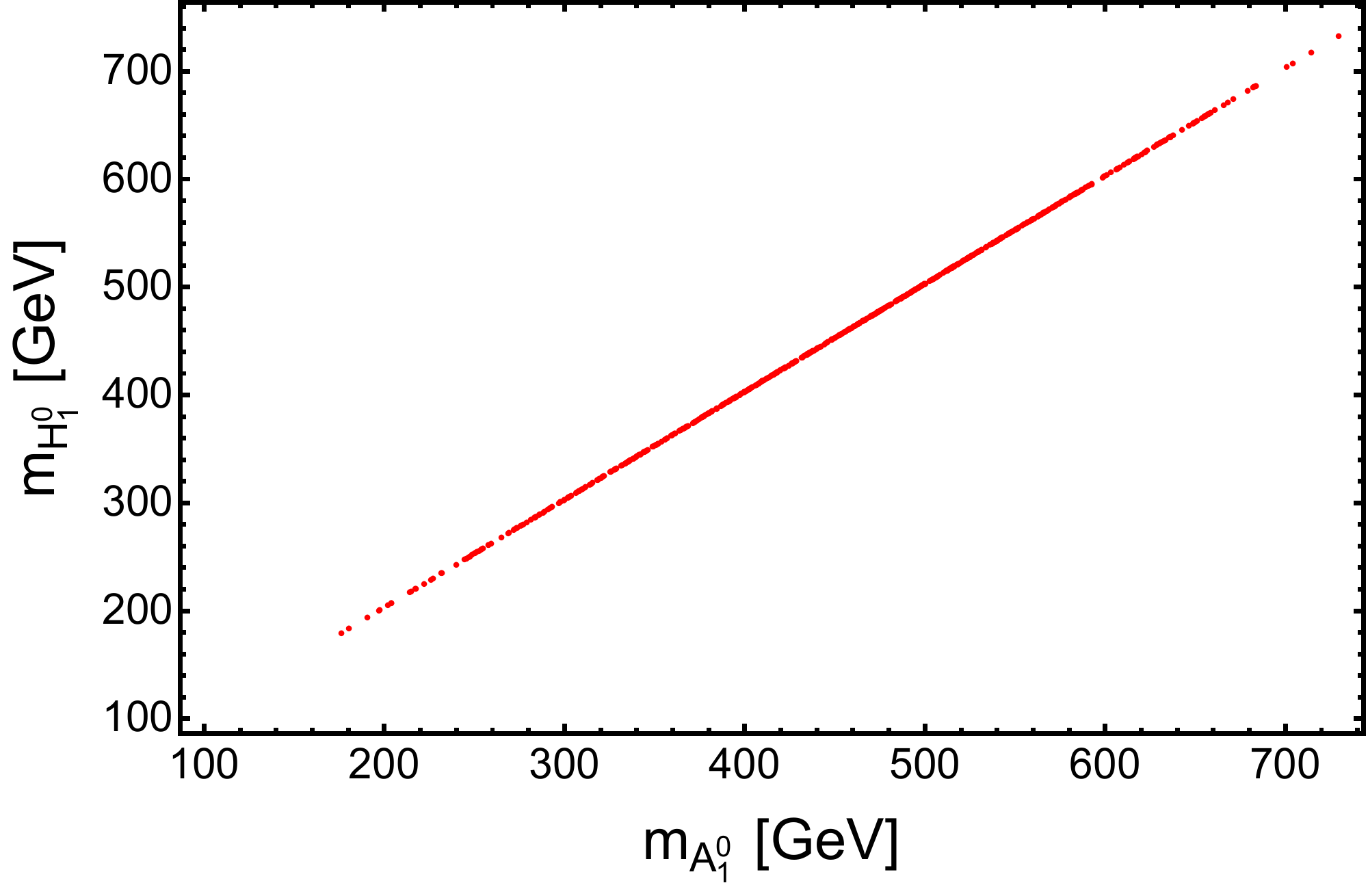}
        \includegraphics[scale=0.27]{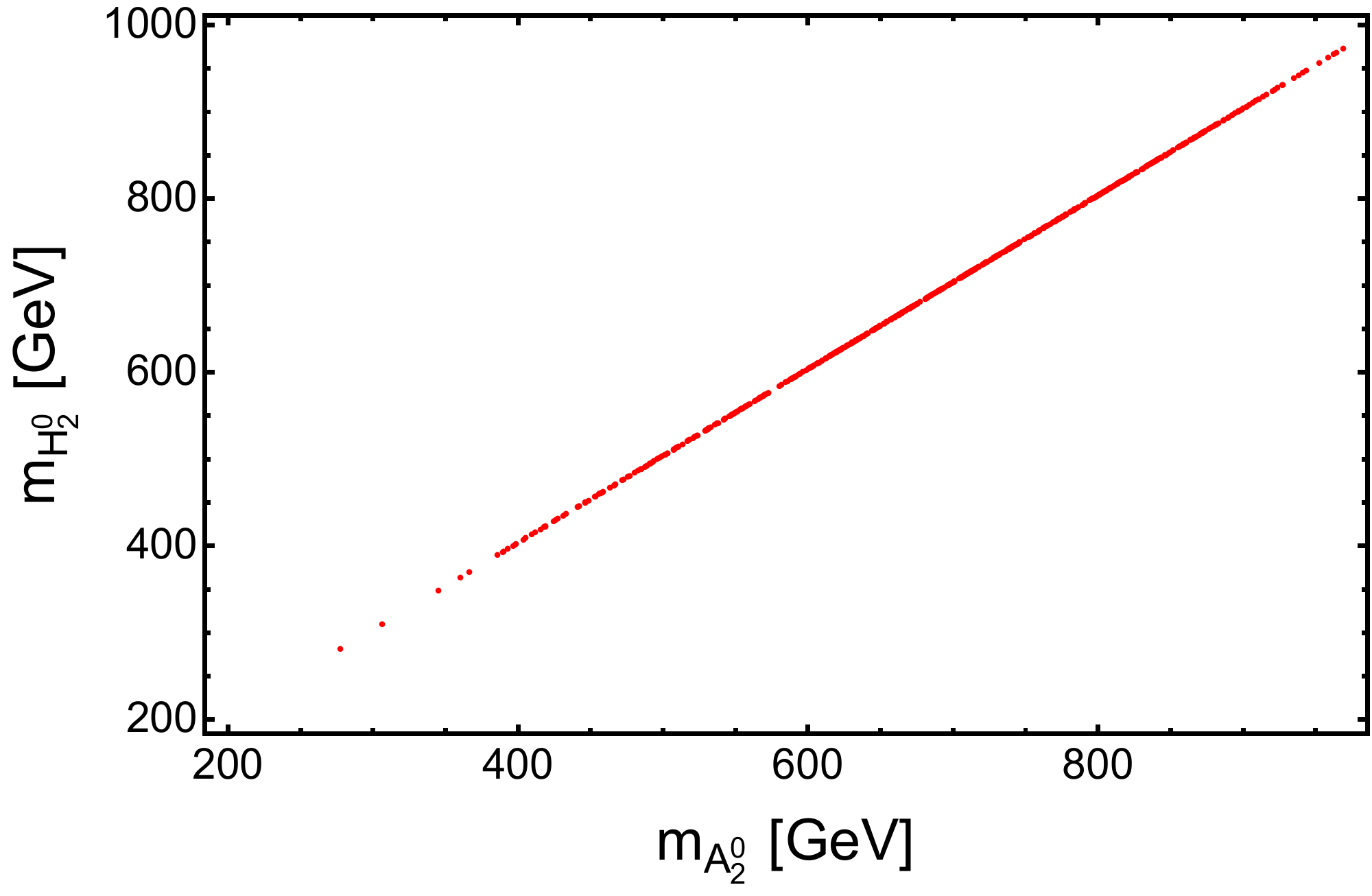}
        \includegraphics[scale=0.27]{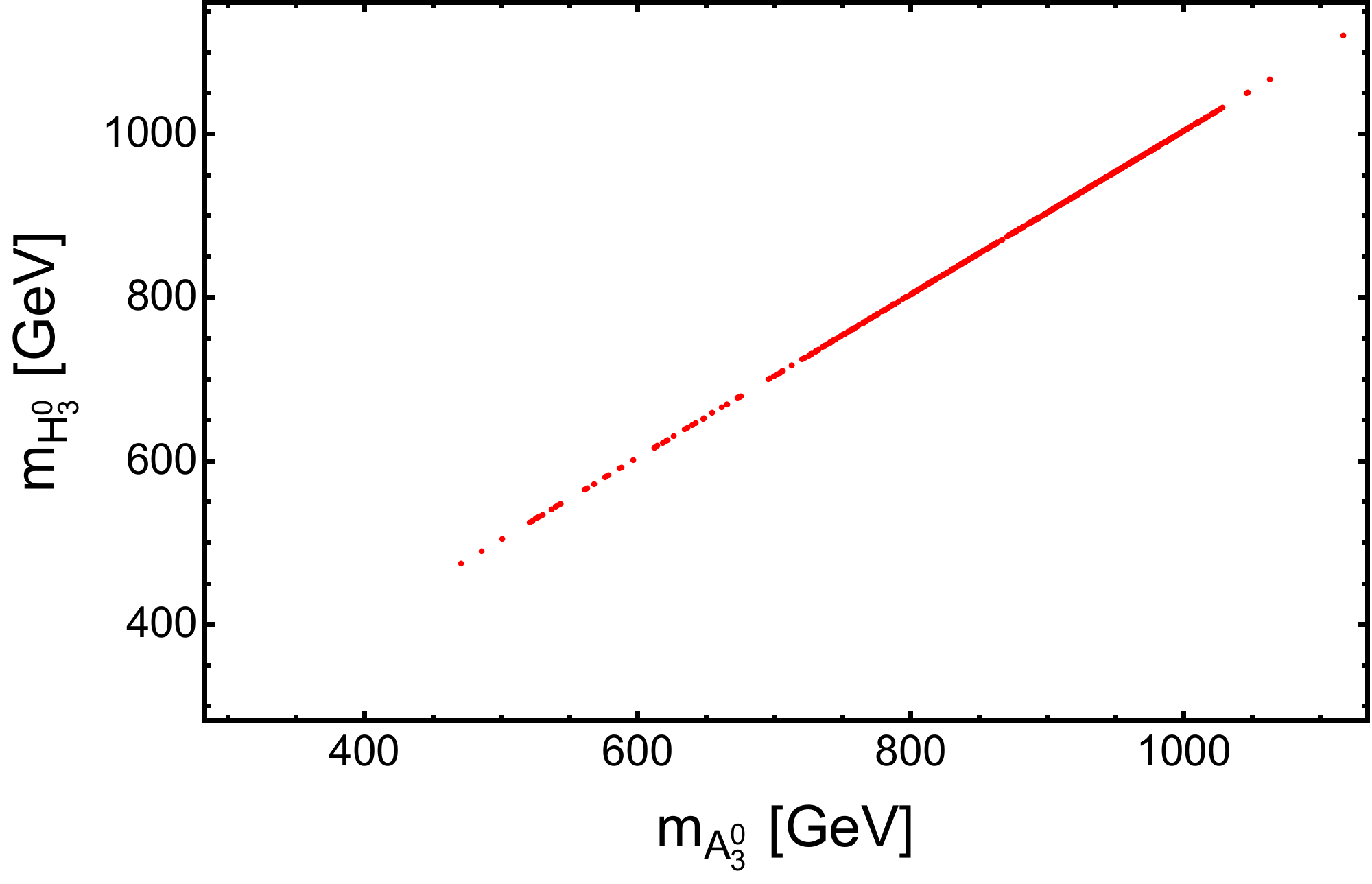}
    \caption{Linear correlations among the neutral and heavy scalar masses.} 
    \label{fig:linear}
\end{figure}

Figures~\ref{fig:correlations} and~\ref{fig:linear} show the correlation among the heavy scalar masses fulfilling all our aforementioned criteria. The linearity that can be seen among them mainly originates from the necessity to satisfy the contributions to the $\rho$-parameter. In all cases, the lightest CP-even neutral scalar is $m_h = 125 \text{ GeV}$, the top quark couples with $\xi^t_h = 1$, and the masses of the CP-even components of the gauge singlet scalars are at the PQ scale, $m_\chi , m_A \sim 10^{12} \text{ GeV}$.  

\section{Dirac neutrinos from PQ symmetry}
\label{app:5higgs}
\noindent
In Sec.~\ref{sec:4HDM}, we chose $\mathcal{X}_{\nu R} = 0$ in order to allow for a simple type-I Majorana seesaw. This is indeed a minimalistic and elegant approach to explain the smallness of neutrino masses without the need to invoke unnaturally small Yukawa parameters. However, here we are precisely exploring a mechanism that gives a Dirac mass to all fermions with natural Yukawas. Moreover, in recent times the topic of Dirac neutrino masses is getting more and more attention \cite{Heeck:2013rpa,Ma:2014qra,Ma:2015mjd,Bonilla:2016zef,Chulia:2016ngi,Ma:2016mwh, Wang:2016lve,Bonilla:2016diq,Hirsch:2017col, Bonilla:2018ynb,Yao:2018ekp,CentellesChulia:2018gwr,Reig:2018mdk,Jana:2019mez,Saad:2019bqf,Jana:2019mgj,Guo:2020qin}, fueled by the as of yet non-observation of neutrinoless double beta decay. It has also been  shown that the PQ symmetry can protect the Dirac nature of neutrinos \cite{Baek:2019wdn, Peinado:2019mrn, delaVega:2020jcp}. Therefore, let us briefly expose how one could extend this mechanism to neutrinos too. We relax the assumption of trivially charged right-handed neutrinos, i.e.\ $\mathcal{X}_{\nu_R} \neq 0$ and add a new Higgs doublet $\phi_\nu$ whose small VEV will give small neutrino masses. Then,  Eqs.~\eqref{eq:neutrinolag} and \eqref{eq:chargephid} get replaced by
\begin{equation}
    -\mathcal{L}_\nu = Y_{\nu, ij} \, \overline{L}_i \, \tilde \phi_\nu \, \nu_{R, j}  \, + \, \text{h.c.},
    \label{eq:diracneutrinolag}
\end{equation}
and
\begin{equation}
\mathcal{X}_{\nu_{R,1}} = \mathcal{X}_{\nu_{R,2}} = \mathcal{X}_{\nu_{R,3}} = \mathcal{X}_{\phi_\nu} + l ,
\label{eq:chargephinu}
\end{equation}
where all the parameters are defined in a straightforward way following the same notation as Eqs.~\eqref{eq:abelian}, \eqref{eq:lagcascade0}, and \eqref{eq:lagcascade}. The same mechanism that gives a hierarchy to the other four Higgs doublets will work here. In order to have this 'cascade mechanism'  to get an appropriate value for the VEV of $\phi_\nu$, we need to include the term $\kappa_{d\nu\chi} (\phi_\nu^\dagger \phi_d) \, \chi$, thus replacing Eq.~\eqref{eq:chargesphi} by
\begin{eqnarray}
\label{eq:diracchargesphi}
&\mathcal{X}_{\chi} = k, \hspace{1cm}  \mathcal{X}_{\phi_d} = n,&\\
&\mathcal{X}_{\phi_\nu} = n-k,\hspace{1cm} \mathcal{X}_{\phi_\mu} = n+k, \hspace{1cm}  \mathcal{X}_{\phi_b} = n+2k, \hspace{1cm}  \mathcal{X}_{\phi_t} = n+3k & ,
\end{eqnarray}
and obtaining a VEV given by
\begin{equation}
    v_\nu \simeq \frac{\kappa_{d\nu\chi} v_\chi v_d}{\mu_\nu^2 + (\lambda_{t\nu 1} + \lambda_{t \nu 2}) v_t^2 + \lambda_{\nu \chi} v_\chi^2} ,
\end{equation}
which can satisfy $v_\nu \sim 0.1$ eV by taking, for example, $v_\chi = 1$ GeV, $v_d = 0.001$ GeV, $\kappa_{d\nu\chi} = 10$ GeV and $\mu^2_\nu = (10^4$ GeV$)^2 \gg v_t^2$. 
Note that now we have one extra degree of freedom for the charges of the model, since we do not  impose  $\mathcal{X}_{\nu_{ R, i}} = 0$. However, this new freedom does not modify $E/N = 8/3$ since neutrinos are gauge singlets.

\bibliography{literature.bib}
\bibliographystyle{apsrev4-1}

\end{document}